\documentclass[aps,prl,graphicx,amsmath,amssymb,reprint]{revtex4-2} 

\usepackage{graphicx}
\usepackage{subfig}
\usepackage{mathtools}

\usepackage{dcolumn}
\newcolumntype{d}[1]{D{.}{\cdot}{#1} }
\usepackage{bm}
\usepackage{xcolor,soul}
\usepackage[normalem]{ulem}

\definecolor{bEpic}{RGB}{24,77,71}
\definecolor{gEpic}{RGB}{150,187,124}
\definecolor{yEpic}{RGB}{250,213,134}
\definecolor{rEpic}{RGB}{198,71,86}
\usepackage{listings}
\lstdefinelanguage{Julia}{
  morekeywords={function, end, return, if, else, elseif, for, while, break, continue, using},
  sensitive=true,
  morecomment=[l]\#,
  morestring=[b]",
}
\usepackage{siunitx}

\begin{document}
\title{Machine learning intermolecular transfer integrals with compact atomic cluster representations}
\date{\today} 



\author{Keerati Keeratikarn}
\affiliation{Department of Physics, Imperial College London, Exhibition Road, London  SW7 2AZ, UK}

\author{Christoph Ortner}
\affiliation{Department of Mathematics, University of British Columbia
Vancouver, BC, Canada V6T 1Z2}

\author{Jarvist Moore Frost}
\affiliation{Department of Chemistry, Imperial College London, 82 Wood Lane, London, W12 0BZ, UK}
\altaffiliation{Department of Physics, Imperial College London, Exhibition Road, London  SW7 2AZ, UK}



\begin{abstract}
Calculating intermolecular charge transfer integrals in organic semiconductors requires substantial computer resource for each individual calculation. 
We might alternatively construct a machine learning model for transfer integrals, which model the full six-degrees of freedom for the relative position of dimer pairs, trained on representative calculations for the molecules of interest. 
Recent developments have produced effective machine learning force fields, which model the total energy of atomic assemblies. 
We extend the Atomic Cluster Expansion (ACE) with the correct symmetries for transfer (kinetic-energy) integrals. 
Combined with a spherical harmonic basis makes, this forms a strong inductive bias and makes for a data efficient model. 
We introduce coarse-grained and heavy-atom representations, and assess the methodology on representative conjugated semiconductors: ethylene, thiophene, and naphthalene.
\end{abstract}




\maketitle

\section{Introduction}\label{introduction}
Organic semiconductors are increasing applied in technical applications of lightweight, flexible, optoelectronics. 
In these materials, the lack of periodicity and relatively low electronic coupling between molecules localises the charge carriers. 
Charge transfer occurs via discrete hopping events between localised states. 
The rate of this hopping is highly dependent on the transfer integral, which is a function of the overlap of the wavefunctions of the molecules, 
\begin{equation}
    \label{equ:1}
    J_{ab} = \langle \psi_a | \hat{H} | \psi_b \rangle,
\end{equation}
where $\psi_a$ and $\psi_b$ denote the frontier molecular orbitals at the respective molecular sites, and $\hat{H}$ represents the electronic Hamiltonian. 
This is the kinetic energy of an electron hopping between sites $a$ and $b$. 
In this work we assume that $a$ and $b$ are discrete but identical molecules. 
In many materials of technical application, the sites can be chromophores which are part of polymers, and can be chemically very different. 
Due to the amorphous or semi-crystalline nature of the material (lack of periodicity) many such couplings exist within a sample. 

The wavefunction on a molecule decreases exponentially with distance, as the electron tunnels into the vacuum. 
Therefore the transfer integral is expected to decrease exponentially with distance. 
The frontier orbitals of a given molecule adopt a complex nodal surface. 
Changing the relative angle of the molecules would therefore expect to pick up a phase factor in the transfer integral: the transfer integral can and should go negative. 
As wavefunctions are required by quantum mechanics to be continuous and smooth, the transfer integral should be smoothly varying. 

\begin{figure*}[t]
\centering
\includegraphics[width=0.6\textwidth]{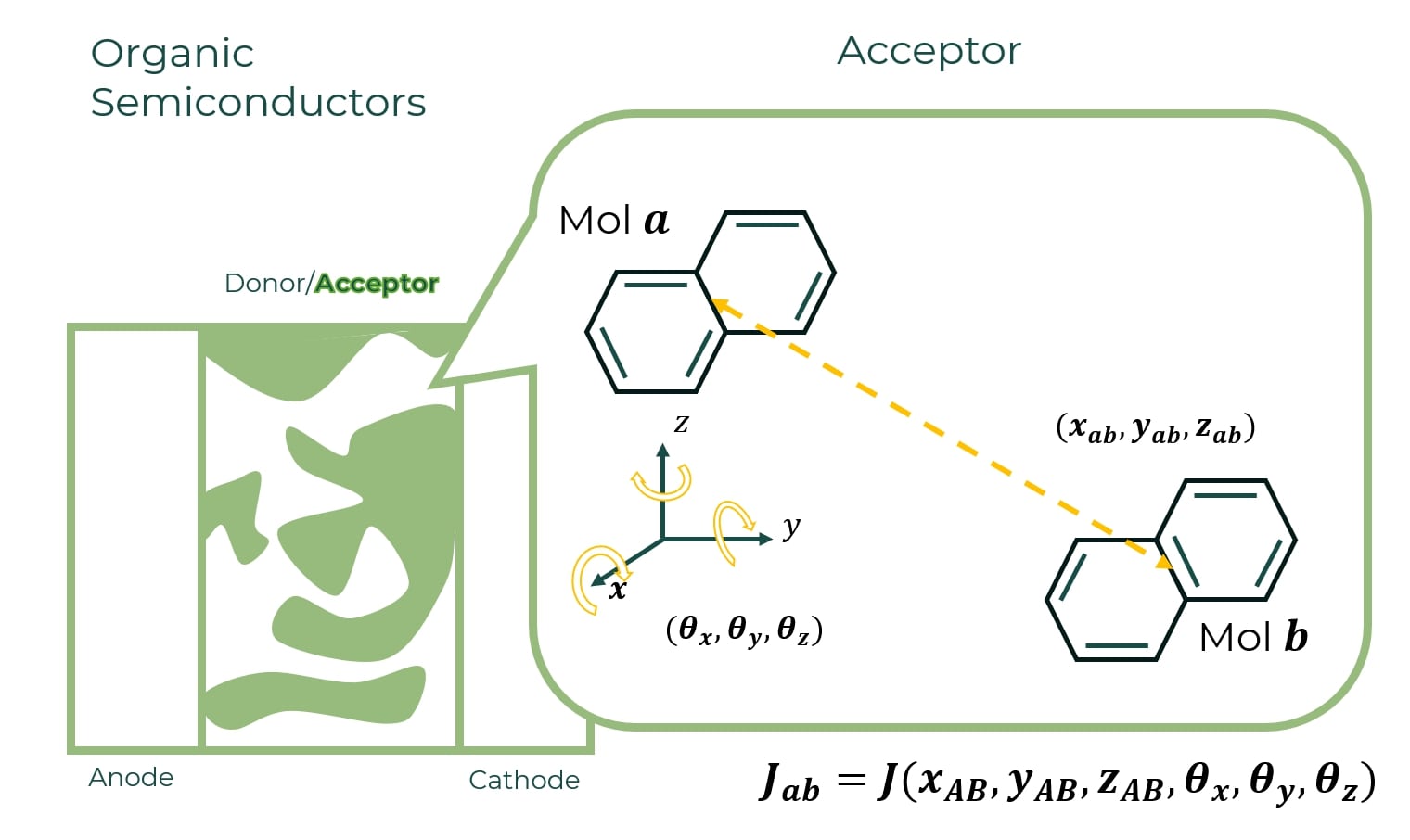}
\caption{\label{fig:1} Schematic representation of charge transfer in organic semiconductors. The active layer features a hybrid donor/acceptor shape situated between the anode and cathode electrodes. 
A magnified perspective illustrates the electronic coupling between two $\pi$-conjugated molecules (Mol $a$ and Mol $b$), defined by their respective centre-of-mass displacement $(x_{ab}, y_{ab}, z_{ab})$ and relative rotational orientations $(\theta_x, \theta_y, \theta_z)$. The transfer integral $J_{ab}$, which dictates the charge transfer rate between molecules, depends on their relative coordinates.}
\end{figure*}

We assume discrete, rigid, molecules. 
Therefore $J_{ab}$ is a 6-dimensional object which we can parameterise with the relative center-of-mass displacement $(x_{ab}, y_{ab}, z_{ab})$ and the relative rotational orientations $(\theta_x, \theta_y, \theta_z)$ between two molecules (see Figure~\ref{fig:1}).

Semi-classical Marcus theory is often used to model charge transfer in these systems, where the microscopic rate of electron transfer between $a$ and $b$ is   
\begin{equation}
    \label{equ:2}
    \Gamma = \frac{2\pi}{\hbar} |J|^2 \frac{1}{\sqrt{4\pi\lambda k_B T}} \exp\left( -\frac{(\Delta G + \lambda)^2}{4\lambda k_B T} \right).
\end{equation}
Here, $\lambda$ denotes the reorganisation energy, $\Delta G$ the Gibbs free energy change, $k_B$ Boltzmann's constant, and $T$ the temperature. 
The rate of charge-transfer is directly proportional to the transfer integral squared. 
This is common to most semi-classical mobility theories, and is inherited from Fermi's Golden Rule. 
For this reason, methods to calculate transfer integrals often do not care for the sign of the transfer integral. 
To deal with quantum-mechanical degeneracies, transfer integrals are often averaged in quadrature, which also loses the sign (phase). 

Calculating transfer integrals poses numerous difficulties. 
Evaluating \ref{equ:1} requires approximations, and a diabatisation step to project the wavefunction of one molecule onto other. 
Accurate methods require a quantum-chemical calculation on the dimer pair to carry out this step. 
Quantum chemical calculations, at the required hybrid density functional theory level or above, are computationally intensive, especially for extensive datasets of molecular dimers \cite{Troisi2011}. 

Transfer integrals exhibit considerable sensitivity to chemical structure \cite{Troisi2006}; minor changes can result in substantial differences in $J$. 
Similarly, for a given chemical structure a slight change in intermolecular orientation can lead to substantial variations.

Therefore the development of effective surrogate models is crucial for facilitating extensive simulations of charge transport in realistic systems \cite{Fratini2016}. 
It is likely that we will always need to fit new models for specific systems, based on quantum-chemical calculations, so we need a reliable and data-efficient method. 
This would enable accurate and high-throughput predictions of charge carrier mobility, and thus iterative computational rational design of high mobility functional materials. 
Similarly, very large assemblies could be simulated, for entire devices and for more complete modelling of charge-transfer biological systems, such as the modelling of photosynthesis. 

In recent years, several machine learning approaches have been devised to forecast transfer integrals.
In 2018, Lederer et al.\cite{Lederer2018} predicted the transfer integrals of the pentacene dimer using a Gaussian Process (GP)\cite{RasmussenW06} model.
The study employed a Gaussian covariance function and a geometric descriptor that incorporates the relative orientations and positions of the molecules inside the system.
In 2020\cite{Rinderle2020}, follow-up work enhanced the model accuracy by using Coulomb matrix (CM)-based descriptors and applying a Gaussian kernel and a Laplacian kernel.

Similarly and separately, in 2019 Wang et al.\cite{Wang2019} estimated the transfer integrals of ethylene dimers using Gaussian Process regression defined by a Gaussian kernel and a Laplacian kernel.
They examined the influence of other CM descriptors—namely intramolecular, intermolecular, and individual-atom contributions—on the predictions.
Their findings indicated that the intermolecular contributions alone are sufficient to adequately characterise the system, as transfer integrals naturally denote quantum-mechanical intermolecular interactions.
In their subsequent study \cite{Wang2020}, they substituted the Gaussian Process regression model with artificial neural networks (ANNs), and saw performance increase.
This enabled the prediction of transfer integrals for more molecular systems, particularly naphthalene dimers in amorphous environments.

In contrast, more recently Hafizi et al. \cite{Hafizi2023} developed a neural-network model that calculates transfer integrals by directly aggregating atomic contributions.
The model effectively represents the complexities of local atomic interactions through the use of atomic local-environment descriptors generated from symmetry functions (SF).
To facilitate the effective selection of training data, they implemented farthest-point sampling (FPS) and active learning methodologies, including query-by-committee (QbC).
This strategy substantially reduces the required size of the training set compared to previous methods.
Furthermore, their research incorporates physics-based models via a $\Delta$-machine learning ($\Delta$-ML) framework, wherein the neural network is trained to amend predictions derived from a pre-existing analytic baseline model.
The integration of data efficiency with physical grounding enhances the model's precision and generalisability.

The Atomic Cluster Expansion (ACE) framework \cite{Drautz2019, Dusson2022, Witt2023} offers a physically informed, robust and systematically augmentable approach for predicting symmetric properties of particle systems. 
ACE provides a symmetry-preserving characterisation of atomic environments, crucial for accurately representing the intricate electronic couplings that influence charge transport. 
The core representation, in terms of spherical harmonics and a radial component, exactly mirrors the mathematical basis used in quantum chemistry, which were in term influenced by the form of atomic orbitals (exact solutions of the Schr\"odinger equation in a $1/r$ potential). 
The direct inclusion higher-order many-body terms with linear computational scaling, along with its versatility in representing both geometric and chemical properties, suggested ACE as potentially being a highly efficient and accurate tool for simulating quantum parameters, including transfer integrals. 
The many-body formalism of ACE has a direct correspondence with the chemistry concept of superexchange, where higher-order effects influence the otherwise pair-wise interaction. 

As a linear model, fitting requires standard and reliable linear algebra techniques. 

\section{Theory}

We build on the \textbf{Atomic Cluster Expansion (ACE)} framework.  Introduced by Drautz~\cite{Drautz2019} in 2019, ACE provides a systematically improvable and efficient many-body descriptor with linear scaling computational cost.


To set the stage for our methodology, we begin by presenting a brief mathematical construction of the \textbf{ACE} model.

\subsection{Linear Atomic Cluster Expansion}
The atomic cluster expansion was original developed to parameterise an energy potential. 
This is a function of the atomic structure which is invariant under permutation of identical atoms as well as under rotation and reflection. We review the framework in this setting before explaining how we apply it to modelling transfer integrals.  

The first step is to decompose the total energy as a sum of atomic energies \cite{Dusson2022, Witt2023, Drautz2019, Musil2021},  
\begin{equation}
\label{equ:3}
E(\left\{ \mathbf{r}_{i}, Z_i \right\}_{i=1}^{J}) 
= \sum_{i=1}^{J} \epsilon(\left\{ \mathbf{r}_{ij}, Z_j, Z_i\right\}_{j \sim i}),
\end{equation}
where $\epsilon$ denotes the atomic energy potential, which depends on the relative positions $\mathbf{r}_{ij} = \mathbf{r}_j - \mathbf{r}_i$ and the chemical elements $Z_i, Z_j$ of its neighbouring atoms, denoted by $j \sim i$. 
For the sake of notational simplicity we define ${\bf x}_{ij} := ({\bf r}_{ij}, Z_j, Z_i)$.

Each site energy $\epsilon$ is then expanded into a hierarchical many-body expansion, often referred to as the \textit{canonical cluster expansion}, centred around atom $i$, and expressed as
\begin{equation}
\label{equ:4}
\begin{split}
    \epsilon(\left\{\mathbf{x}_{ij}\right\}_{j \sim i}) =&
 V^{(0)} +
\sum_{{j}_{1}} V^{(1)}(\mathbf{x}_{i{j}_{1}}) \\ 
& \qquad \,\, +
\sum_{{j}_{1}<{j}_{2}} V^{(2)}(\mathbf{x}_{i{j}_{1}},\mathbf{x}_{i{j}_{2}}) \\
&+\cdots+ \hspace{-1em}
\sum_{{j}_{1}<{j}_{2}<\cdots<{j}_{N}} \hspace{-1em} V^{(N)}(\mathbf{x}_{i{j}_{1}},\cdots,\mathbf{x}_{i{j}_{N}}),
\end{split}
\end{equation}
where $V^{(N)}$ represents the potential energy contribution from an $(N+1)$-atom cluster, including the central atom and $N$ of its neighbours \cite{Dusson2022, Ho2024}. This formulation results in prohibitive combinatorial computational complexity. 

To achieve a computationally efficient cluster expansion model, Drautz \cite{Drautz2019} modified the restricted summations in the \textit{canonical} cluster expansion~\eqref{equ:4} to an unrestricted summation, 
\begin{equation}
\label{equ:5}
\begin{split}
   \epsilon(\left\{\mathbf{x}_{ij}\right\}_{j\sim i}) =&
 U^{(0)} +
\sum_{{j}_{1}} U^{(1)}(\mathbf{x}_{i{j}_{1}}) \\ 
& \qquad \,\, +
\sum_{{j}_{1}{j}_{2}} U^{(2)}(\mathbf{x}_{i{j}_{1}},\mathbf{x}_{i{j}_{2}}) \\
&+\cdots+
\sum_{{j}_{1}{j}_{2}\cdots{j}_{N}}U^{(N)}(\mathbf{x}_{i{j}_{1}},\cdots,\mathbf{x}_{i{j}_{N}}).
\end{split}
\end{equation}
This cluster expansion with these unrestricted summations is called the \textit{self-interacting cluster expansion} due to the fact that the summation now includes non-physical clusters with repeated atoms. 

Each interaction potential $U^{(N)}$ is parameterised in terms of tensor products of \textbf{single particle basis functions} $\phi_{k}(\mathbf{x}_{ij})$,
\begin{equation}
\label{equ:6}
\begin{split}
\sum_{{j}_{1}{j}_{2}\cdots{j}_{N}}&U^{(N)}(\mathbf{x}_{i{j}_{1}},\cdots,\mathbf{x}_{i{j}_{N}})=\\
&\sum_{{j}_{1}{j}_{2}\cdots{j}_{N}}
\sum_{{k}_{1}\cdots{k}_{N}}
\tilde{c}^{(Z_i)}_{{k}_{1}\cdots{k}_{N}}\prod_{t=1}^{N}\phi_{k_{t}}(\mathbf{x}_{i{j}_{t}}),
\end{split}
\end{equation}
where $\tilde{c}^{(Z_i)}_{{k}_{1}\cdots{k}_{N}}$ is the $(N+1)$-body's expansion coefficient and each $k_{t}$ denotes a multi-index of quantum numbers, $k_t = (\zeta_t, n_{t}, l_{t}, m_{t})$, which we explain below when we define the one-particle basis.

Exchanging the summation over $j_1, \dots, j_N$ with the product results in the computationally efficient formulation~\cite{Drautz2019,Witt2023,Dusson2022}, 
\begin{align}
    \label{equ:7}
    A_{k}^{(i)} 
    &= \sum_{j}\phi_{k}(\mathbf{x}_{ij}), \\ 
    \label{equ:8}
    \textbf{A}^{(i)}_{\mathbf{k}}
    &= \prod_{t=1}^{N_{\bf k}} A_{k_{t}}^{(i)} \\ 
    \label{equ:78-epsilon}
    \epsilon(\left\{\mathbf{x}_{ij}\right\}_{j\sim i}) 
    &= 
    \sum_{{\bf k} \in \mathcal{K}} 
        \tilde{c}_{\bf k}^{(Z_i)} \textbf{A}^{(i)}_{\mathbf{k}},
\end{align}
where $N_{\bf k}$ denotes the ``length'' of the multi-index ${\bf k}$ and $\mathcal{K}$ the set of selected multi-indices corresponding to the set of selected features. 


The formulation \eqref{equ:78-epsilon} already incorporates permutation invariance due to the pooling operation in \eqref{equ:8}. To incorporate also reflection and rotational invariance, one selects a one-particle basis of the form 
\begin{equation}
\label{equ:10}
\begin{aligned}
    \phi_{\zeta nlm}(\mathbf{x}_{ij}) 
    &= \phi_{\zeta nlm}({\bf r}_{ij}, Z_j, Z_i) \\
    &= R_{\zeta n}(r_{ij}, Z_j, Z_i) Y_l^{m}(\hat{\mathbf{r}}_{ij}),
\end{aligned}
\end{equation}
where $R_{\zeta n}$ is a radial basis (to be specified) and $Y_l^m$ are the spherical harmonics, for which a convenient representation of $O(3)$ is available. This enables a linear transformation from ${\bf A}^{(i)}_{\bf k}$ features to $O(3)$-invariant features, 
\begin{equation}
\label{equ:11}
\begin{split}
    \textbf{B}_{{\bm \zeta}\mathbf{nl}\mu}^{(i)}
    =
    \sum_{{\bf m}} \mathcal{U}^{{\bm \zeta}\textbf{nl}\mu}_{\textbf{m}} 
\textbf{A}^{(i)}_{{\bm \zeta}\mathbf{nlm}},
\end{split}
\end{equation}
where $\mathcal{U}^{{\bm \zeta}\textbf{nl}\mu}_{\textbf{m}}$ are generalized Clebsch-Gordan coefficients and the new index $\mu$ enumerates all possible invariant couplings for a given $({\bm \zeta}{\bf nl})$ tuple. Details of this construction can be found in \cite{Drautz2019,Witt2023,Dusson2022}. 


The resulting expansion of of the atomic energy potential $\epsilon$ in terms of the \textbf{symmetrised many body basis} is given by 
\begin{equation}
\label{equ:12}
\begin{split}
    \epsilon(\left\{\mathbf{x}_{ij}\right\}_{j\sim i}) =& 
    \sum_{({\bm\zeta}{\bf nl}\mu) \in \mathcal{F}}  
    c^{(Z_i)}_{{\bm\zeta}{\bf nl}\mu} \textbf{B}_{{\bm \zeta}\mathbf{nl}\mu}^{(i)},
\end{split}
\end{equation}
where $\mathcal{F}$ is the set of symmetrised features selected for the expansion. 
Recall from \eqref{equ:3} that the total energy of the system is obtained by summing over all atomic energies.

\subsection{Application to transfer integrals}
In the present work, our objective is to fit the transfer integrals \( J_{ab} \) between molecules $a$ and $b$ using the Atomic Cluster Expansion (ACE) model. 

Analogous to equation~\eqref{equ:12}, we can express the \textit{site} transfer integral by solving for the parameter \(\tilde{c}\) as
\begin{equation}
\label{equ:13}
\mathcal{J}_{ab}\!\left(\{\mathbf{x}_{ij}\}_{j \sim i}\right) = 
\sum_{({\bm\zeta},\mathbf{n},l,\mu) \in \mathcal{F}}  
\tilde{c}^{(Z_i)}_{{\bm\zeta}\mathbf{n}l\mu} 
\mathbf{B}^{(i)}_{{\bm \zeta}\mathbf{n}l\mu,ab},
\end{equation}
where the chemical species \(Z_i\) and the index \({\bm\zeta}\) need not be the true atomic species but can be chosen to be a \textit{pseudo-atomic species} (any categorical variable describing an abstract particle), as discussed in the \textbf{Data Analysis and ACE Basis Construction} section.

After determining the parameters $\tilde{\bm c}$ for the model in~\eqref{equ:13}, the total transfer integral of the molecular dimer system is then constructed as a summation over the \textit{site} contribution 
\begin{equation}
\label{equ:14}
J_{ab}\!\left(\{\mathbf{x}_{i}\}_{i=1}^{J}\right) 
= \sum_{i=1}^{J} 
\mathcal{J}_{ab}\!\left(\{\mathbf{x}_{ij}\}_{j \sim i}\right).
\end{equation}

In formulating the model~\eqref{equ:13} and~\eqref{equ:14} for the transfer integral, we ensure the predicted \( J_{ab} \) preserves the key physical symmetries and locality of the underlying electronic Hamiltonian.  
The transfer integrals between molecular orbitals \(\psi_a\) and \(\psi_b\) is defined as
Eq.~\eqref{equ:1} which, by the Hermiticity of the Hamiltonian \(\hat{H} = \hat{H}^\dagger\), satisfies
\begin{equation}
\label{equ:15}
J_{ab} = J_{ba}^*.
\end{equation}
For systems with real-valued molecular orbitals, this reduces to \(J_{ab} = J_{ba}\), indicating that the transfer integral is a \textit{symmetric scalar quantity} under the exchange of monomers \(a\) and \(b\).  
This property is naturally preserved in the ACE representation through symmetric parameterisation of the coefficients \(\tilde{c}^{(Z_i)}_{{\bm\zeta}\mathbf{n}l\mu,ab}\) or through training on symmetrised data, ensuring that
\begin{equation}
\label{equ:16}
\mathcal{J}_{ab}\!\left(\{\mathbf{x}_{ij}\}_{j \sim i}\right)
= \mathcal{J}_{ba}\!\left(\{\mathbf{x}_{ij}\}_{j \sim i}\right).
\end{equation}

Since the transfer integrals depend only on the \textit{relative geometry} of the molecular dimer and not on its absolute position or orientation, the ACE basis functions  
\(\mathbf{B}^{(i)}_{{\bm\zeta}\mathbf{n}l\mu}\) are constructed to be rotationally and translationally invariant, guaranteeing that
\begin{equation}
\label{equ:17}
J_{ab}\!\left(\{\mathbf{x}_{i}\}_{i=1}^{J}\right)
= J_{ab}\!\left(\{\mathbf{x}_{i}'\}_{i=1}^{J}\right),
\quad \text{for } 
\mathbf{x}_i' = \hat{\mathbf{Q}}\mathbf{x}_i + \mathbf{t},
\end{equation}
where \(\hat{\mathbf{Q}}\) is any rotation matrix and \(\mathbf{t}\) a translation vector.  
This invariance ensures that \(J_{ab}\) depends on the internal geometry of the dimer, consistent with physical symmetry requirements.

The spatial decomposition in equation~\eqref{equ:10} reflects the \textit{locality} of electronic interactions.  
In quantum mechanical terms, the total coupling can be written as a sum of local orbital--orbital interactions,
\begin{equation}
\label{equ:18}
J_{ab} = \sum_{i \in a}\sum_{j \in b} 
\langle \phi_i | \hat{H} | \phi_j \rangle,
\end{equation}
where each term decays rapidly with interatomic distance due to the localised nature of atomic orbitals.  
By analogy, each site contribution \(\mathcal{J}_{ab}(\{\mathbf{x}_{ij}\})\) in the ACE model depends only on neighboring atoms within a finite cut-off radius, ensuring that \(J_{ab}\) is expressed as a sum of local, symmetry-preserving contributions:
\begin{equation}
\label{equ:19}
\mathcal{J}_{ab}\!\left(\{\mathbf{x}_{ij}\}_{j \sim i}\right)
= \mathcal{J}_{ab}\!\left(\{\mathbf{x}_{ij}\}_{|\mathbf{x}_{ij}| < r_\mathrm{cut}}\right).
\end{equation}

Hence, the representation in equations~\eqref{equ:11}--\eqref{equ:12} guarantees that the fitted transfer integrals \(J_{ab}\) possess the correct Hermitian, rotational, and permutational symmetries, while admitting a spatially local decomposition consistent with the physical structure of intermolecular electronic coupling.

\begin{figure*}[t]
\centering
\includegraphics[width=1.0\textwidth]{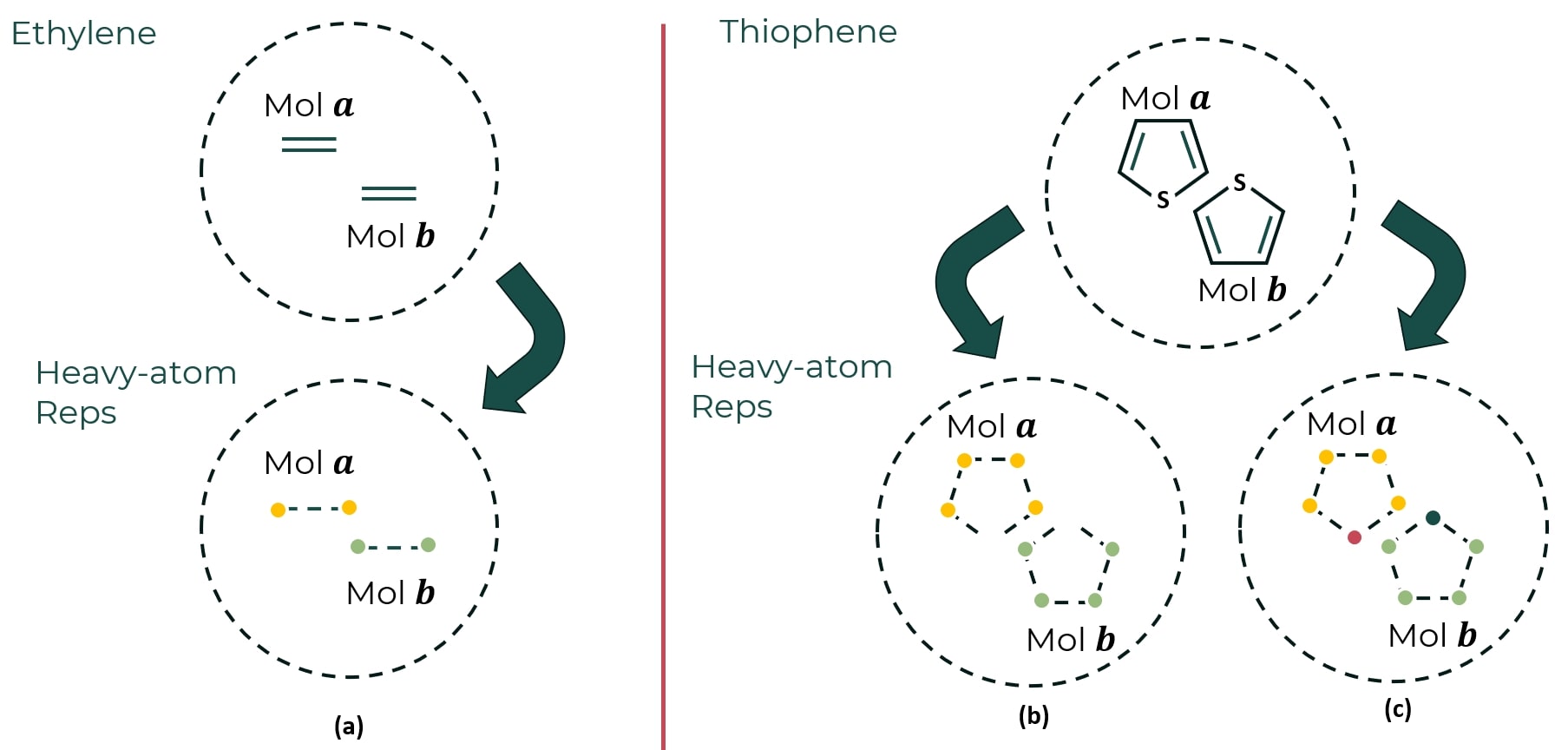}
\caption{\label{fig:2} Illustration of the \textit{Heavy-atom} representations used for ACE descriptor construction for (left) \textit{ethylene} and (right) \textit{thiophene} dimers.
 (a) The ethylene dimer is represented using only carbon atoms (yellow and green dots for Mol~$a$ and Mol~$b$, respectively). Panels (b) and (c) illustrate variations of the \textit{Heavy-atom} representation: (b) includes only carbon atoms, while (c) incorporates both carbon and sulphur atoms as distinct elements.}
\end{figure*}

\subsection{ACE Hyperparameters}
There are two key hyperparameters that must be selected to complete the specification of a linear ACE model: the radial basis $R_{\zeta n}$ and the feature set $\mathcal{F}$. 


We employ the standard framework for constructing a radial basis proposed by Witt et al. \cite{Witt2023}, 
\begin{equation}
\label{equ:20}
\begin{split}
& R_{\zeta n}(r_{ij}, Z_j, Z_i) \\ 
&\quad = f_{\text{env}}(r_{ij}, Z_j, \zeta) \cdot P_n\big( y(r_{ij}, Z_{j}, \zeta)\big) \cdot \delta_{Z_i \zeta},     
\end{split}
\end{equation}
where $f_{\text{env}}$ is an envelope function encoding element-dependent decay behavior, $P_{n}(y)$  is an orthogonal polynomial basis in a transformed coordinate $y_{ij} = y(r_{ij}, Z_j, Z_i)$ (note that $\zeta = Z_i$ when $\delta_{Z_i \zeta} \neq 0$), and term $\delta_{Z_i \zeta}$ denotes a one-hot embedding of the centre element. 
We choose Legendre polynomials for $P_n$, with the transformed coordinate $y(r_{ij}, z_{j}, \zeta)$ capturing the element-specific distance mapping.

For our coordinate transform we take the rational function 
\begin{equation}
    \label{eq:coord_trans_y}
    y_{ij} 
    = y(r_{ij}, Z_j, Z_i) 
    = \bigg(1 + a \frac{(r/r_0)^q}{1 + (r/r_0)^{q-p}} \bigg)^{-1},
\end{equation}
where $r_0 = r_0(Z_j, Z_i)$ is a hyperparameter specifying where to concentrate numerical resolution, $a$ is chosen to maximize the gradient of $y$ at $r_{ij} = r_0$, and $p, q$ are further hyperparameters specifying the asymptotic behavious of $y_{ij}$ as $r_{ij} \to 0, \infty$. Given the specification of the coordinate transform, we can define $y_{\rm cut}^{(ij)} := y(r_{\rm cut}, Z_j, Z_i)$ and then define the envelope function
\begin{equation}
    f_{\rm env}(r_{ij}, Z_j, Z_i) 
    = (y_{ij}-1)^2 (y_{ij} - y_{\rm cut}^{(ij)})^2,
\end{equation}
which ensures a smooth transition to zero as $r_{ij} \to 0, r_{\rm cut}$. 
The selection of the three remaining hyperparameters $r_0, p, q$ is problem-dependent. 
We found this critical for the performance of the linear ACE model, and will be discussed in detail below.

It remains to select the feature set $\mathcal{F}$. The rationale for the coordinate transformation $y_{ij} = y(r_{ij}, Z_j, Z_i)$ is that the atomic potential $\epsilon$ will be much smoother in these new coordinates than in the original $r_{ij}$ coordinates, which then enables a sparse basis selection. 
Concretely, we specify a maximum correlation order $N$ and a maximum total degree $D_{\rm tot}$ and define 
\begin{equation}
    \mathcal{F} := \big\{ 
      ({\bm \zeta}{\bf nl}q) \, \big| \,
      N_{\bf nl} \leq N, 
      {\textstyle \sum_{t}} n_t + l_t \leq D_{\rm tot} \big\}.
\end{equation}
This given two additional hyperparameters, $N, D_{\rm tot}$ to be selected. 
 

\begin{figure*}[t]
\centering
\includegraphics[width=0.8\textwidth]{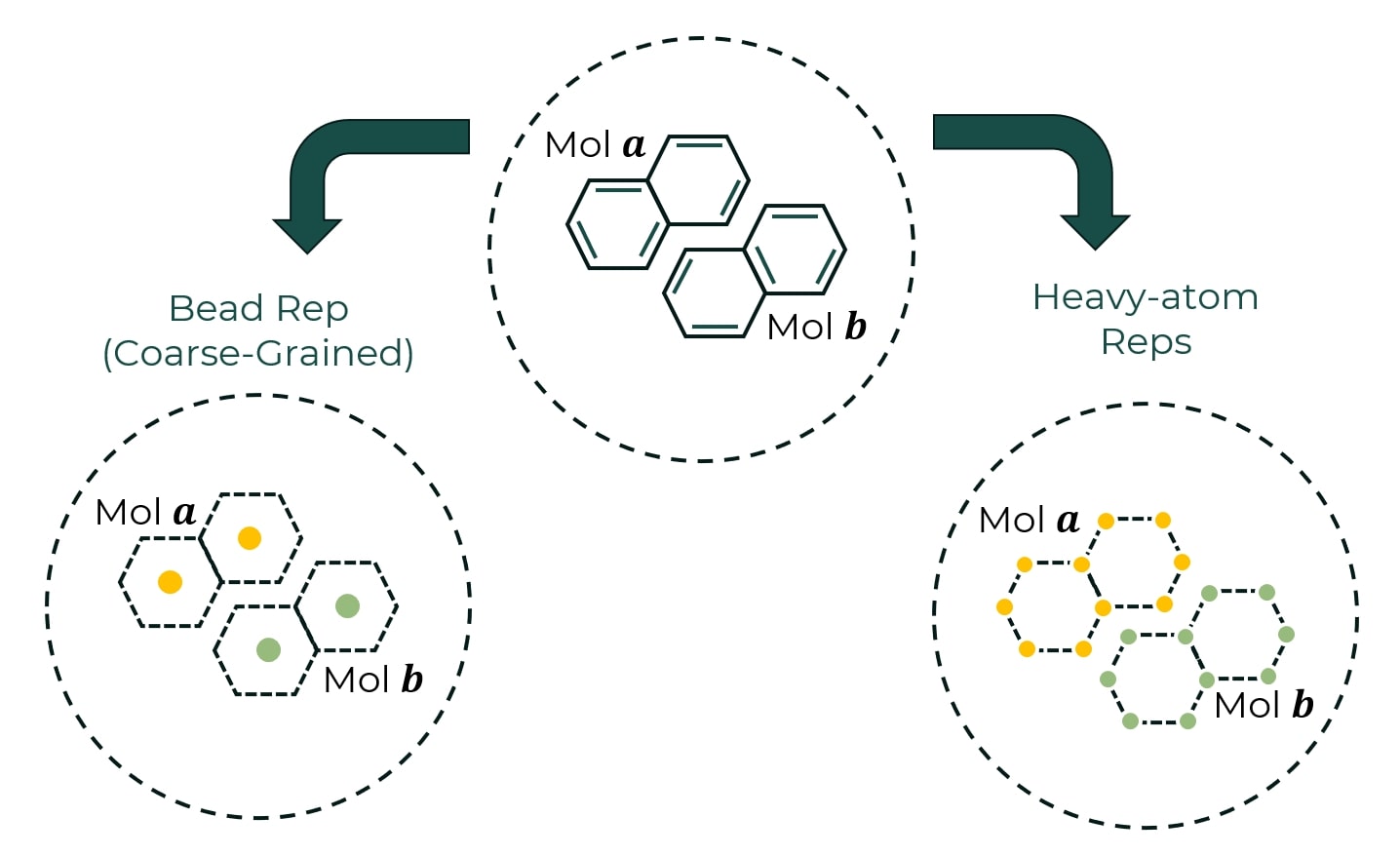}
\caption{\label{fig:3} Illustration of two types of molecular representations used for constructing ACE descriptors. The central diagram shows a dimer of \textit{Naphthalene} (Mol~$a$ and Mol~$b$). The left pathway depicts the \textit{Bead} (\textit{coarse-grained}) representation, in which each aromatic ring is represented by a single Bead. The right pathway illustrates the \textit{Heavy-atom} Representation, where only heavy (Heavy-atom) atoms are included. }
\end{figure*}

\section{Data preparation and analysis}

\subsection{Dataset generation and dimer representation}

In this work we are developing a new method. 
Therefore we need a diverse dataset, with the correct qualitative behaviour, but with little care for the quantitative predictions. 
To this end we use relatively lightweight calculations of transfer-integrals from tight-binding and semi-empirical (ZINDO) based electronic structure methods. 
These lightweight methods enabled us to easily generate large datasets, and experiment with different sampling approaches while developing our models. 

\begin{figure*}[t]
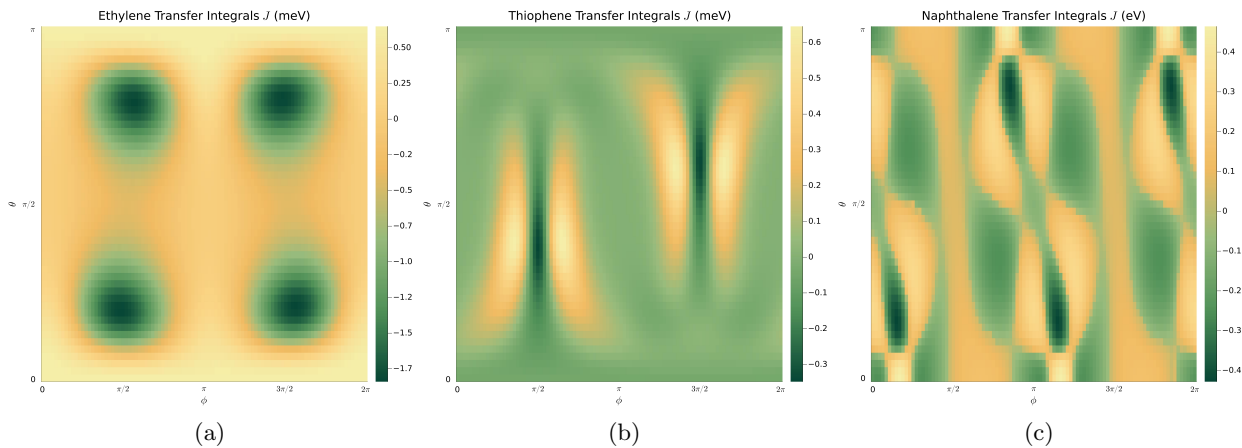
\centering
\captionsetup{justification=centering}
    \begin{tabular}{ccc}
        \includegraphics[width=0.3\linewidth,height=\textheight,keepaspectratio]{Figures/Ethylene/new_ethylene_J_heatmap.jpg} &
        \includegraphics[width=0.3\linewidth,height=\textheight,keepaspectratio]{Figures/Thiophene/new_Thiophene_J_heatmap.jpg} &
        \includegraphics[width=0.3\linewidth,height=\textheight,keepaspectratio]{Figures/Naphthalene/new_Naphthalene_J_heatmap.jpg} \\
        (a) & (b) & (c) \\[6pt]
    \end{tabular}
    \caption 
    {\label{fig:heatmap}
    Heatmaps shows the 2-dimensional transfer integrals of ethylene (a), thiophene (b), and naphthalene (c).}
\end{figure*}

We consider three different molecular dimers: \textit{ethylene}, \textit{thiophene}, and \textit{naphthalene}. Each dataset is constructed to systematically sample relevant geometric configurations as follows:

\begin{itemize}
    \item \textbf{\textit{ethylene} dataset:} 
    The simplest possible conjugated electronic material is ethylene ($C_2 H_4$). 
    There is no technical application (ethylene is a gas; the bandgap is beyond the optical; the ionisation potential is so large that we cannot inject holes or electrons). 
    Nonetheless, it has a well defined transfer integral. 
    
    This dataset was generated using the open-source \textsc{xtb-dipro} package~\cite{Kohn2023} which we edited to pass through the calculated phase (negative) transfer integrals.
    We sampled a three-dimensional surface,   
    \[
    J(x_{ab}, y_{ab}, z_{ab}) := J(r_{ab}, \theta_{ab}, \phi_{ab}),
    \]
    where the intermolecular separation \( r_{ab} \in [6.25,\; 7.00]~\text{\AA} \), the polar angle \( \theta_{ab} \in [0,\; \pi] \), and the azimuthal angle \( \phi_{ab} \in [0,\; 2\pi] \). The relative orientation of molecule \( B \) is fixed at \( \theta_{z} = \pi/6 \) with respect to molecule \( A \).

    The example of ethylene's J is illustrated in Fig.~\ref{fig:heatmap}(a). The transfer inteegrals are calculated where one of the ethylenes is orientated with respect to another at \( r_{ab} = 6.25~\text{\AA}\).
    
    \item \textbf{\textit{Thiophene} dataset:} 
    Most modern organic semiconductors contain 'hetero' atoms taking part in the frontier orbitals beyond carbon. 
    Thiophene ($SC_4H_4$) is the most simple model for a heteroatom containing conjugated electronic material. 
    
    This dataset was generated analogously to the ethylene dataset. The only modification is in the intermolecular distance, with \( r_{ab} \in [7.25,\; 7.75]~\text{\AA} \), while the angular degrees of freedom and fixed orientation \( \theta_{z} = \pi/6 \) are preserved.

    Figure~\ref{fig:heatmap}(b) shows the case of thiophene's J. The transfer integrals are computed with one thiophene orientated relative to another at \( r_{ab} = 7.50~\text{\AA}\).
    
    \item \textbf{\textit{Naphthalene} dataset:} 
    Napthalene is an organic electronic material composed of two fused benzene rings. 
    We used this as a model for much larger fused ring structures, and in the development of a coarse-grained representation. 
    Molecular Orbital Overlap (MOO) is an extreme efficient approach to calculate transfer integrals which exploit the orthogonal nature of the intermediate neglect of differential overlap Hamiltonian. 
    However, this Hamiltonian is not accurate for the push-pull (donor-acceptor) chemistry of more complex modern organic electronic materials. 
    This dataset was generated using the \textsc{pyMOO} package~\cite{Kirkpatrick2007}. 
    We uniformly sampled the two-dimensional angular surface of the transfer integrals:  
    \[
    J(x_{ab}, y_{ab}, z_{ab}) := J(\theta_{ab}, \phi_{ab}),
    \]
    with a fixed intermolecular distance \( r_{ab} = 12.00~\text{\AA} \). The angular variables \( \theta_{ab} \) and \( \phi_{ab} \) vary over the surface of the unit sphere, and the relative orientation of molecule \( B \) is fixed at \( \theta_{z} = \pi/3 \) with respect to molecule \( A \).

    This example of naphthalene transfer integrals is shown in Figure~\ref{fig:heatmap}(c). At \( r_{ab} = 12.0~\text{\AA}\), the transfer integrals are calculated with one naphthalene orientated with respect to another.
\end{itemize}

Before constructing the ACE descriptors $\textbf{B}_{{\bm \zeta}\mathbf{nl}\mu}^{(i)}$ in Equation~\eqref{equ:11}, we introduce an efficient representation of the molecules to reduce the dimensionality of the descriptor space while retaining the essential geometric and electronic features specific to transfer integral prediction.

Instead of employing a standard all-atom neighbour representations, we propose two alternative schemes: (i) a \textit{Heavy-Atom} representation, and (ii) a \textit{Bead} (\textit{Coarse-Grained}) representation.

This approach is motivated by the physical origin of transfer integrals, which arise from couplings between the frontier molecular orbitals. 
Accordingly, it is reasonable to either (i) focus selectively on the atomic orbitals localised on heavy atoms, such as carbon, which play the dominant role in composing the frontier orbitals and thus the transfer integral or (ii) group atomic-orbital-based contributions into effective molecular orbitals via a coarse-grained representation for individual conjugated units.
 
Since the transfer integral is an intermolecular property, we found it important to distinguish between \textit{sites} belonging to different molecules.
We illustrate the \textit{Heavy-atom} approach in Figure~\ref{fig:2} for \textit{ethylene} and \textit{Thiophene} dimers, and on Figure~\ref{fig:3} (right) for the \textit{Naphthalene} dimer.
We assign distinct labels to each molecule, reusing the element specificity of force-fields: the yellow atoms in Mol~$a$ are treated as pseudo-\textbf{C} elements, while the green atoms in Mol~$b$ are labelled as pseudo-\textbf{Si} elements. 
This labelling enforces a separation of molecular environments and ensures that intra- and intermolecular interactions are treated distinctly within the ACE framework.

\begin{figure*}[t]
\centering
\includegraphics[width=0.8\textwidth]{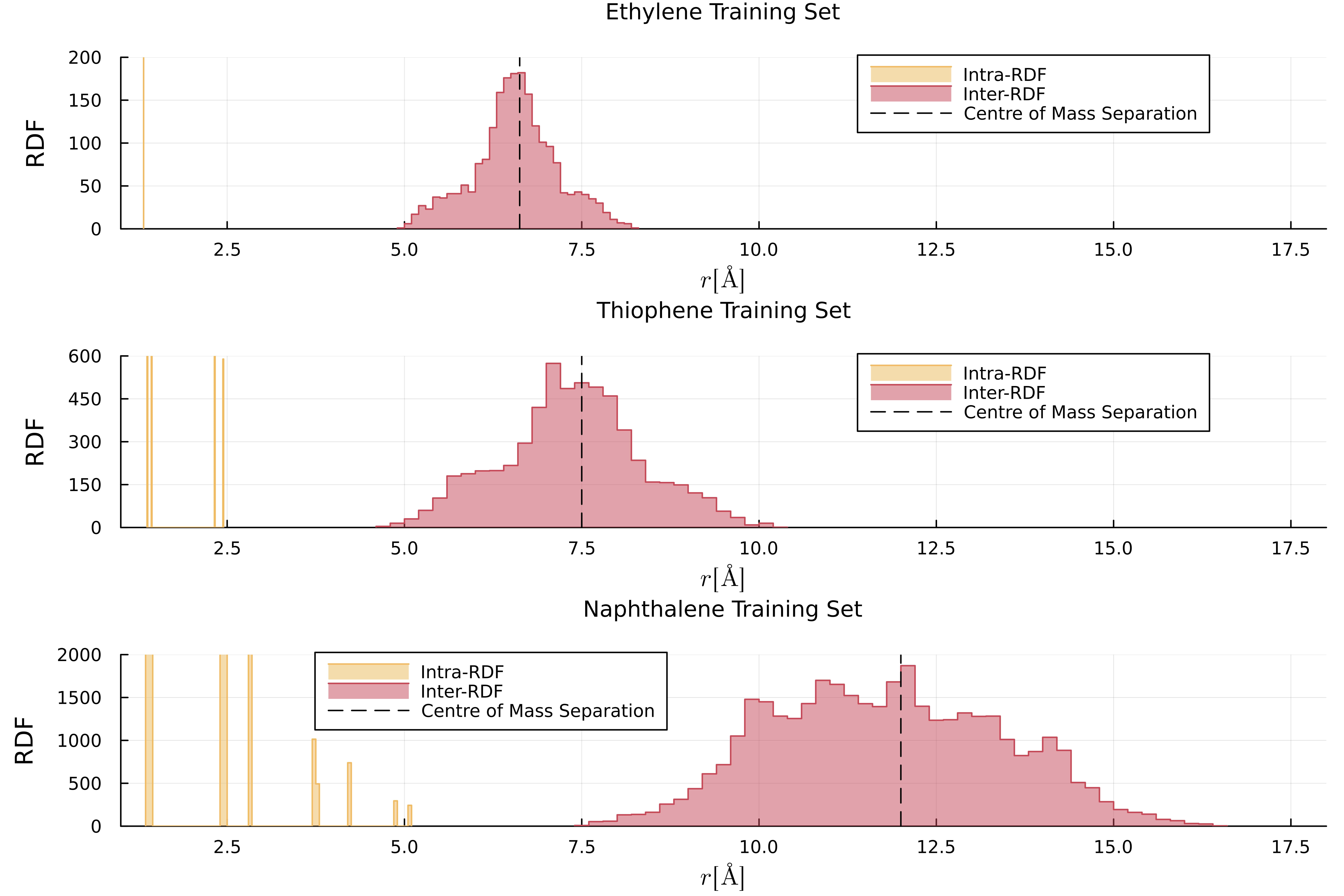}
\caption{\label{fig:4}  Analysis of \textit{ethylene} (upper), \textit{thiophene} (middle), and \textit{naphthalene} (lower) dimer RDFs for the ACE descriptor construction for the \textit{Heavy-atom} representation.}
\end{figure*}

\subsection{Data analysis and ACE basis construction}

Our next step in constructing a successful model involved analyzing the intra- and intermolecular radial distribution functions (RDFs). 
The carbon-carbon pairwise distance we plot in Figure~\ref{fig:4}: \textit{ethylene} (top), \textit{thiophene} (middle), and \textit{naphthalene} (bottom). 
The yellow histogram represents the intramolecular RDF, while the red histogram corresponds to the distribution of intermolecular pairwise distances. 
The dashed vertical lines indicate the fixed centre-of-mass separation of 6.63, 7.50, and 12.00~\AA{ employed in the construction of our dimer configurations.

For \textit{naphthalene}, we used both \textit{Bead} and \textit{Coarse-Grained} representations. 
The organic electronic community has regularly used such coarse-grained (and even entirely empirical) models for transfer integrals, and they could make a natural representation for considering a polymer-polymer transfer integral, or more complex fused ring molecules. 
Wang et al.~\cite{https://doi.org/10.48550/arxiv.2502.04661} recently showed that coarse-grained ACE force-field models are useful for molecular dynamics. 
A similar analysis can be performed to visualise the RDF of the \textit{naphthalene} dimer configuration in the \textit{Bead} representation. The RDF is illustrated in the upper panel of Figure~\ref{fig:5}. The analysis of the RDF guides the construction of radial basis functions (RBFs). 

We specify the hyperparameters which control the structure and behaviour of the basis set as following: 

\begin{figure*}[t]
\centering
\includegraphics[width=0.8\textwidth]{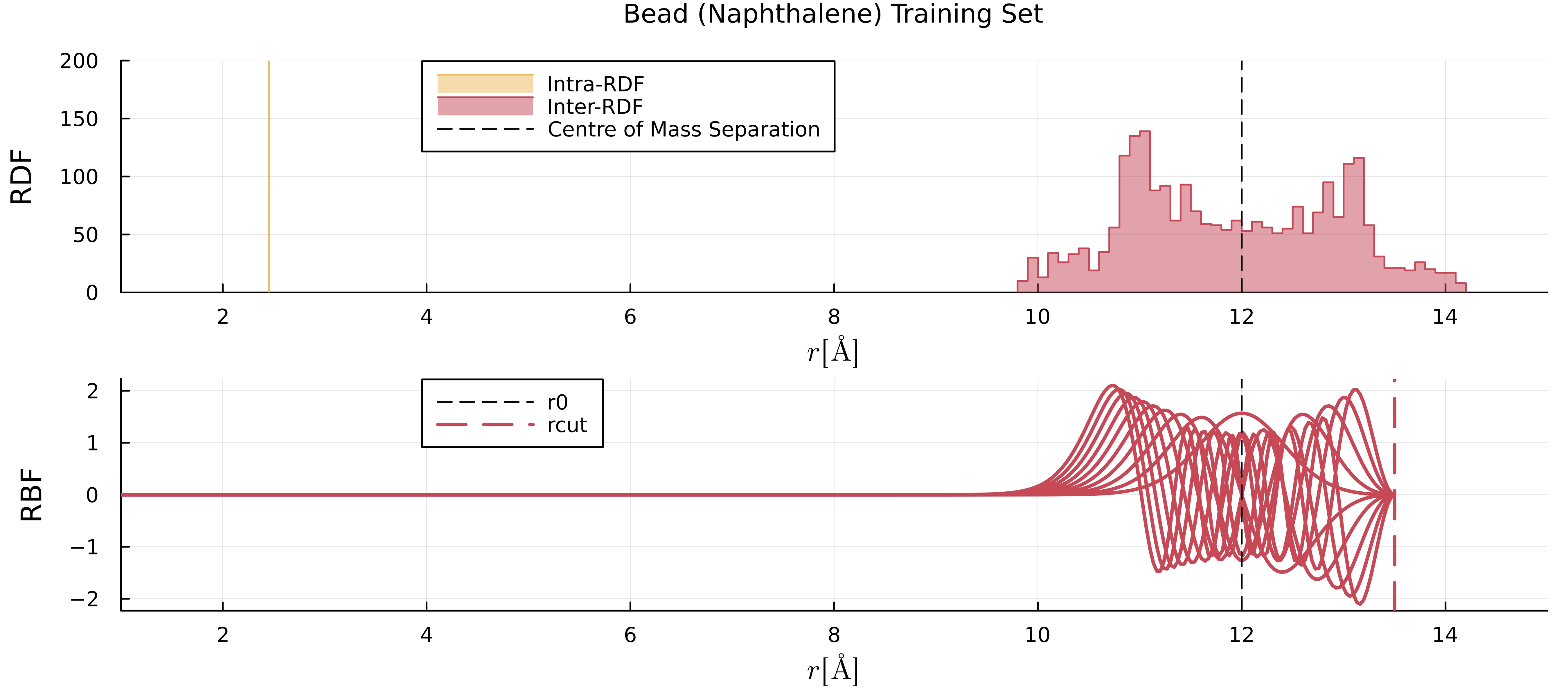}
\caption{\label{fig:5}  Analysis of naphthalene dimer RDFs and RBFs  used in the ACE descriptor construction for the \textit{coarse-grained (Bead)} representation.}
\end{figure*}

\begin{itemize}
    \item 
    Element parameter specifies the chemical species. In this work, it is chosen to be \textbf{C} and \textbf{Si} for almost all the cases, since we treat the atoms (\textit{Beads}) in those dimers as pseudo-C and pseudo-Si elements. With the exception of Pathway \textbf{(c)} in the Figure~\ref{fig:2}, we set the parameter to be \textbf{C}, \textbf{Si}, \textbf{O} and \textbf{S} in order to support the four-element system.
    \item 
    The parameter $N$ sets the interaction order $N$ (corresponding to $N+1$-body correlations). The default setting is $N=2$ corresponding to the 3-body correlation. Testing the model with $N=2$ to 5 does not yield a significant improvement in accuracy compared to the substantial increase in computational cost associated with evaluating the ACE basis at higher body orders.
    \item 
    The total degree parameter $D_{\rm tot}$ constrains the maximum total polynomial degree of the basis functions. We use $D_{\rm tot}=10$ as the lowest degrees of the total polynomial to be able to fit the transfer integrals. The model was tested with $D_{\rm tot} =5$ to 30, and we recommend not increasing $D_{\rm tot}$ beyond 15, as higher values provide diminishing returns in fitting accuracy relative to the computational cost.
    \item 
    The parameters $r_{0}$ and $r_{\text{cut}}$ respectively define an (approximate) equilibrium bond length, specific to the chemical species, and the cut-off radius beyond which basis functions vanish smoothly.
    These two hyperparameters can be chosen from the analysis of the radial distribution of the training datasets shown in Figures~\ref{fig:4} and~\ref{fig:5}. For instance, we selected $r_{0} = 7.0,\;7.5,\;12.0$~\AA~and $r_{\text{cut}} = 7.5,\;9.0,\;13.6$~\AA~for ethylene, thiophene, and naphthalene dimers
    %
    %
    %
    \item 
    The pair transform function corresponds to the transformation defined as $y(r_{ij}, Z_j, \zeta)$ in Equation~\eqref{equ:20}, typically using the Agnesi function, which enhances resolution near the nearest-neighbour distance and improves the accuracy of the interatomic potential. 
    
    The defaults in ACE models optimised for interatomic energy are $p = 1$ and $q = 4$. We set $p$ and $q$ extremely high, because they help to compress the basis functions (close to $r_0$) in order to be adapted to the relevant intermolecular RDF support ranges. 
    
    The parameter $q$ controls the shape of the left hand side of the radial basis functions (RBFs). Smaller values of $q$ extend the RBFs toward shorter interatomic distances, while larger values contract them closer to the equilibrium bond length $r_0$. 
    
    The parameter $p$ governs the sharpness and localisation of the envelope near $r_0$, particularly on the right hand side. A smaller $p$ results in a more gradual rise of the envelope function near $r_{\text{cut}}$, whereas a larger $p$ produces a steeper transition and stronger localisation around $r_0$. We recommend setting $q$ within the range of 41--43 and $p$ within 17--23.
    \item 
    Reference energy $E_{\text{ref}}$ defines the one-body (atomic reference) energy. By setting $E_{\text{ref}}=0$ for each Bead type, we enforce the physical constraint that the transfer integral vanishes for isolated molecules or Beads.
\end{itemize}

With all hyperparameters thus defined, the ACE model radial basis is constructed as shown in the lower panel of Figure~\ref{fig:5}. The upper panel shows the intra- and intermolecular RDFs as a function of bead-pairwise distance, with a dashed line indicating the fixed centre-of-mass separation of 12~\AA. The lower panel depicts the corresponding RBFs used to capture only intermolecular environments.

\section{Results and discussion}

\begin{figure*}[t]\centering
\captionsetup{justification=centering}

    \begin{tabular}{cccc}
        \includegraphics[width=0.225\linewidth,height=\textheight,keepaspectratio]{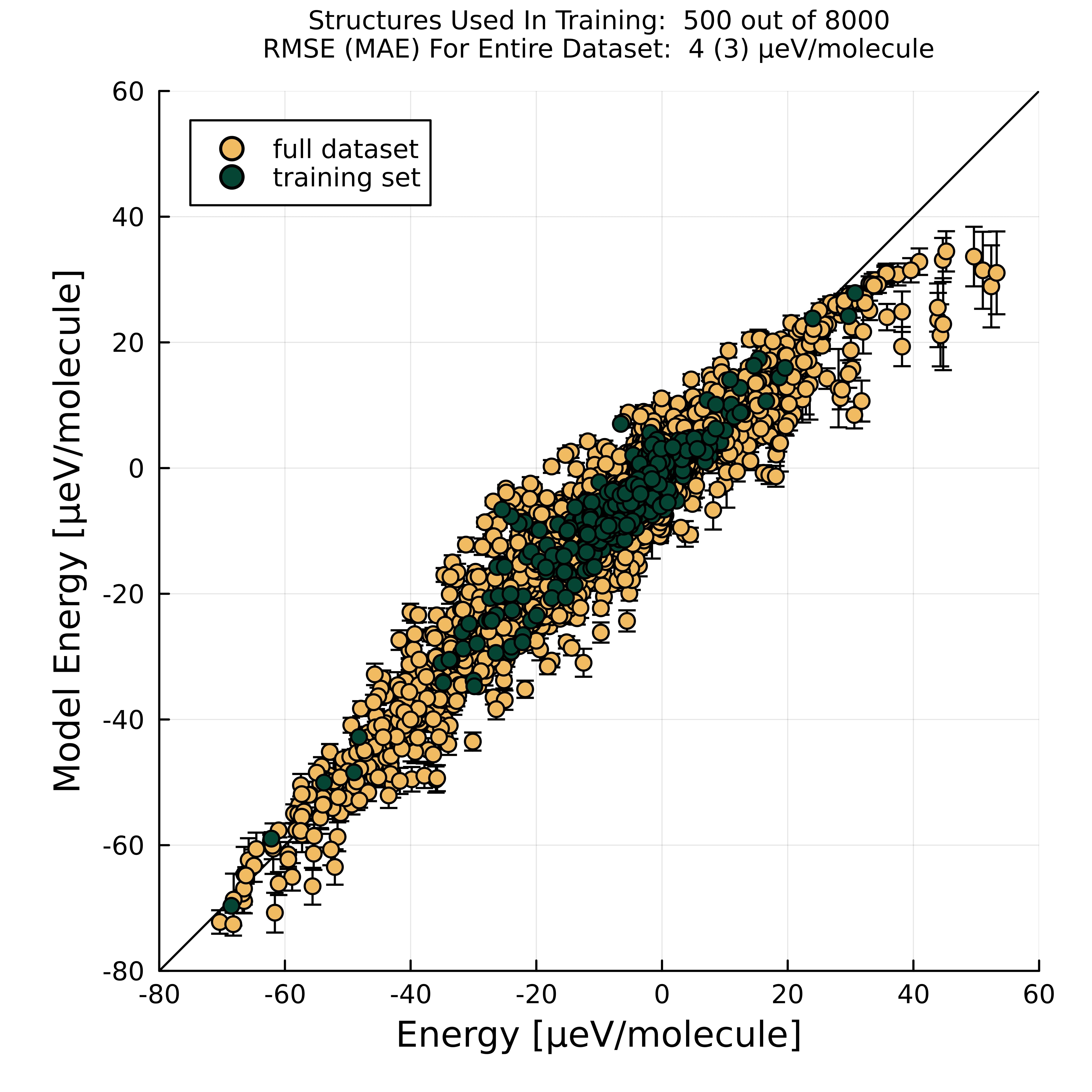} &
        \includegraphics[width=0.225\linewidth,height=\textheight,keepaspectratio]{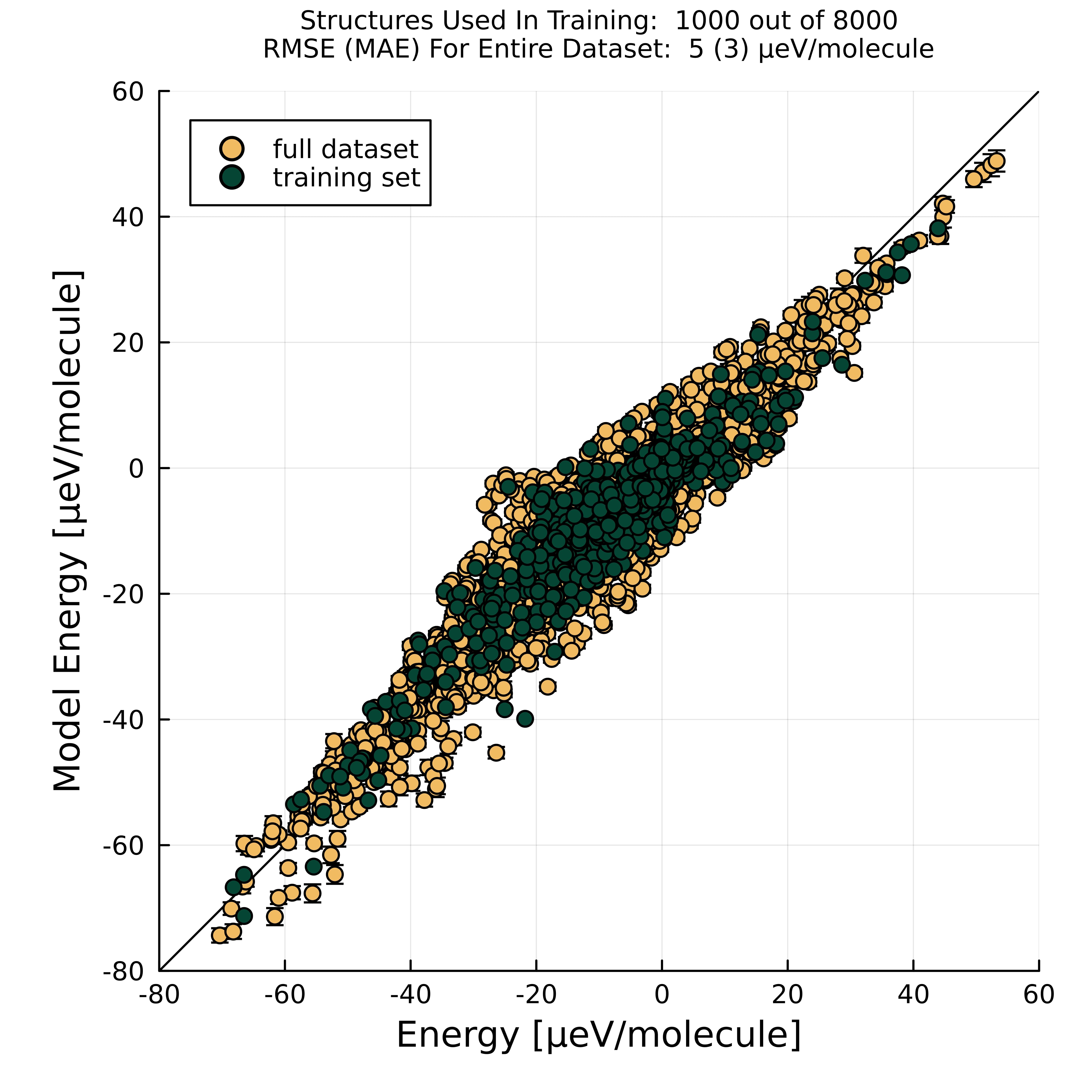} &
        \includegraphics[width=0.225\linewidth,height=\textheight,keepaspectratio]{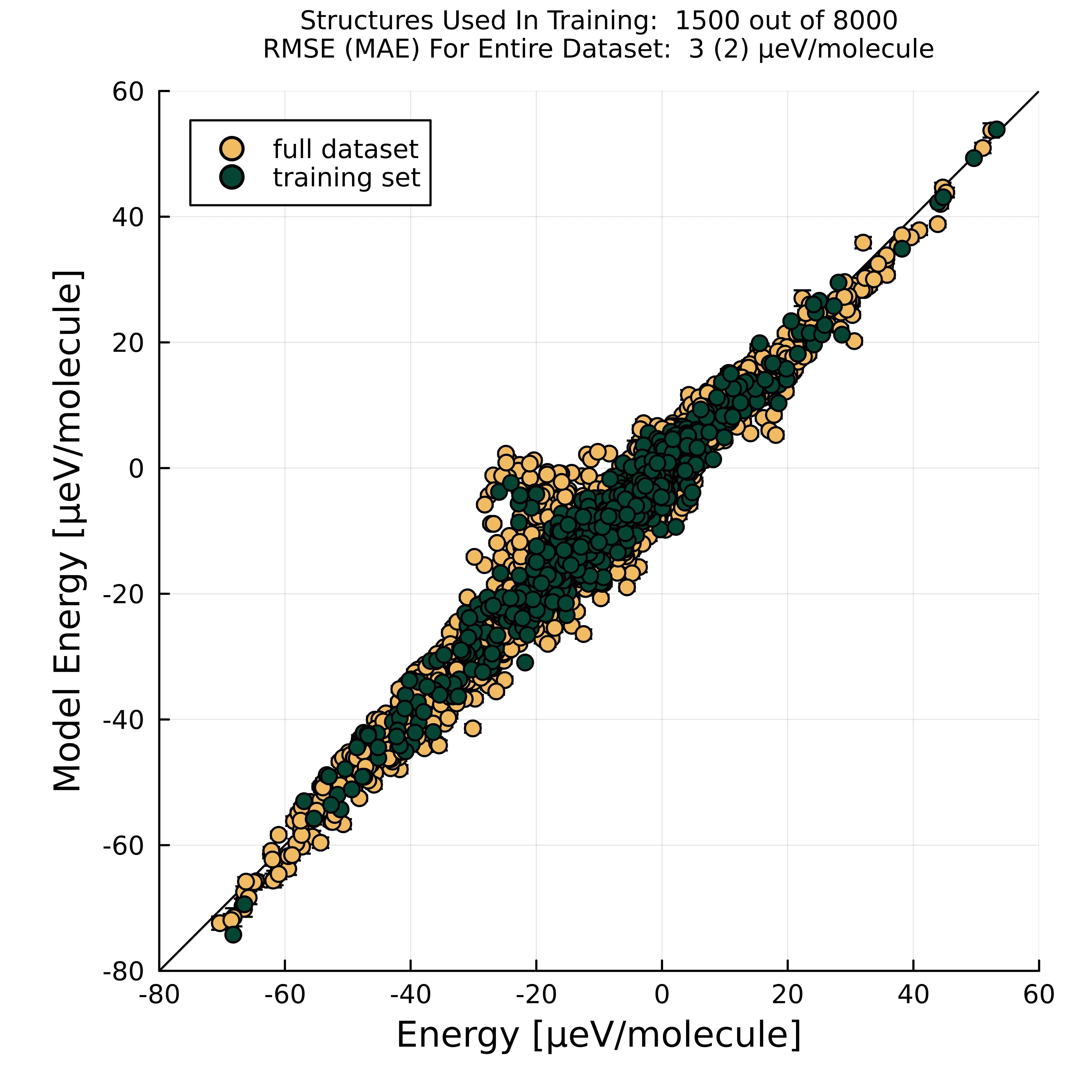} &
        \includegraphics[width=0.225\linewidth,height=\textheight,keepaspectratio]{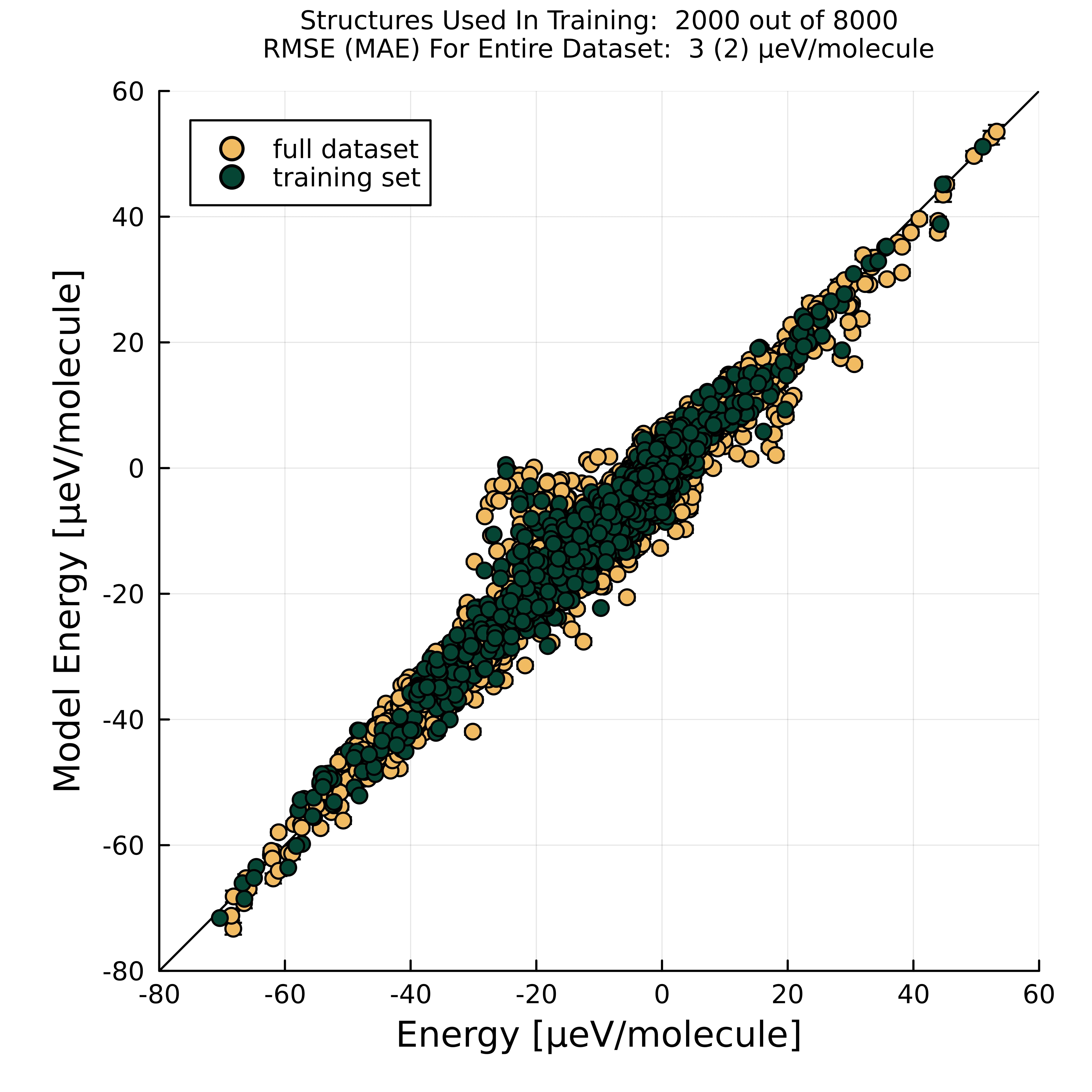}  \\
        (a) & (b) & (c) & (d)\\[6pt]
    \end{tabular}
    \begin{tabular}{cccc}
        \includegraphics[width=0.225\linewidth,height=\textheight,keepaspectratio]{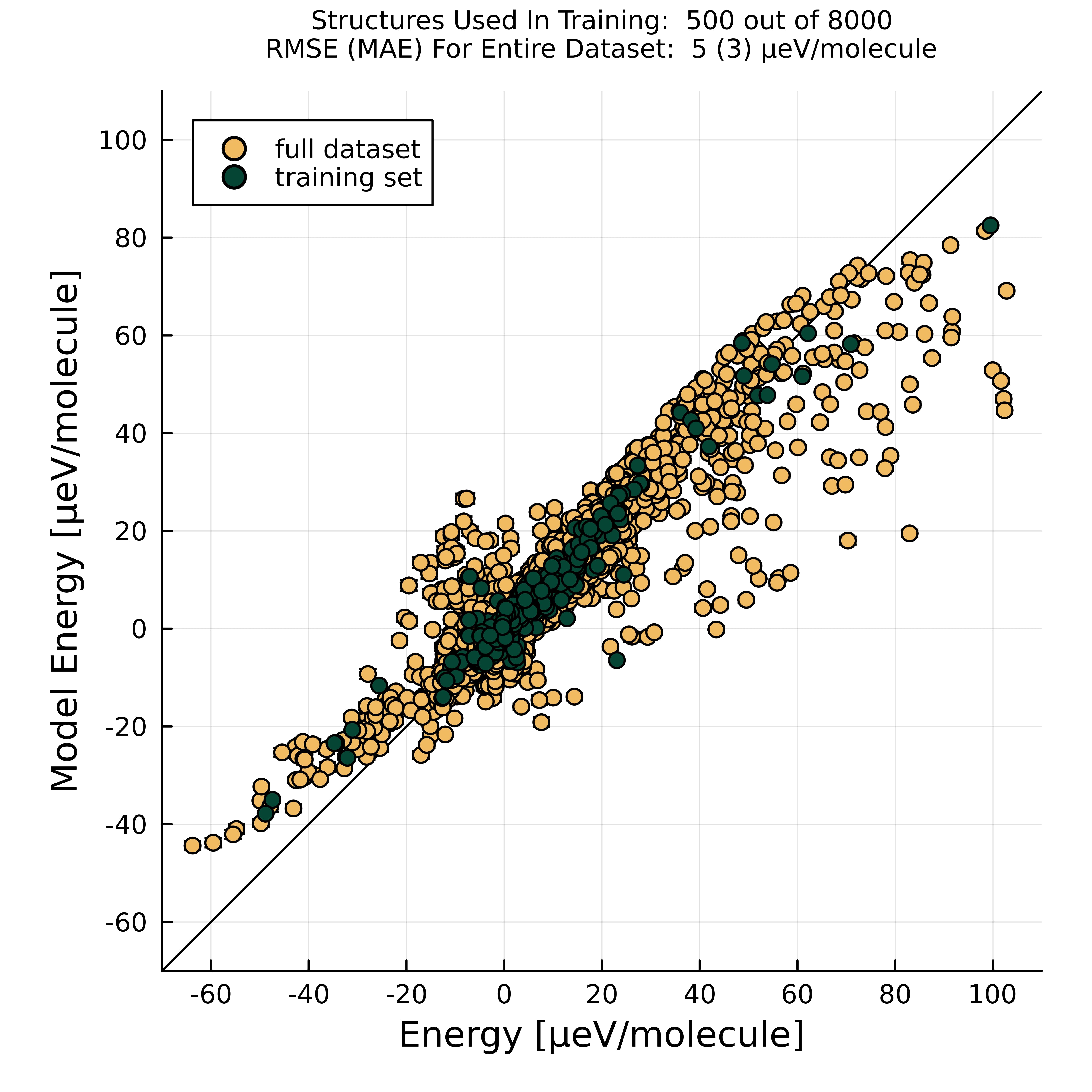} &
        \includegraphics[width=0.225\linewidth,height=\textheight,keepaspectratio]{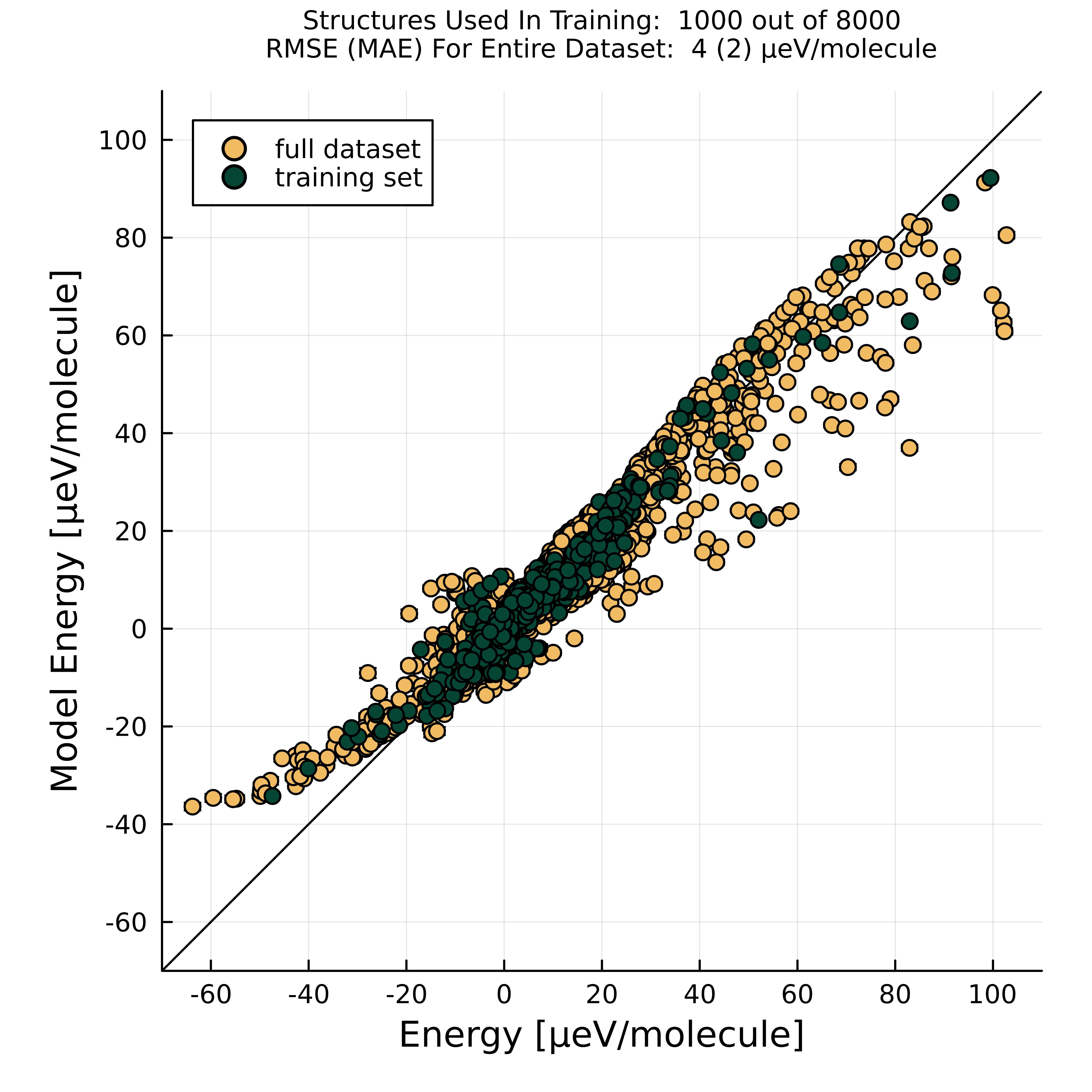} &
        \includegraphics[width=0.225\linewidth,height=\textheight,keepaspectratio]{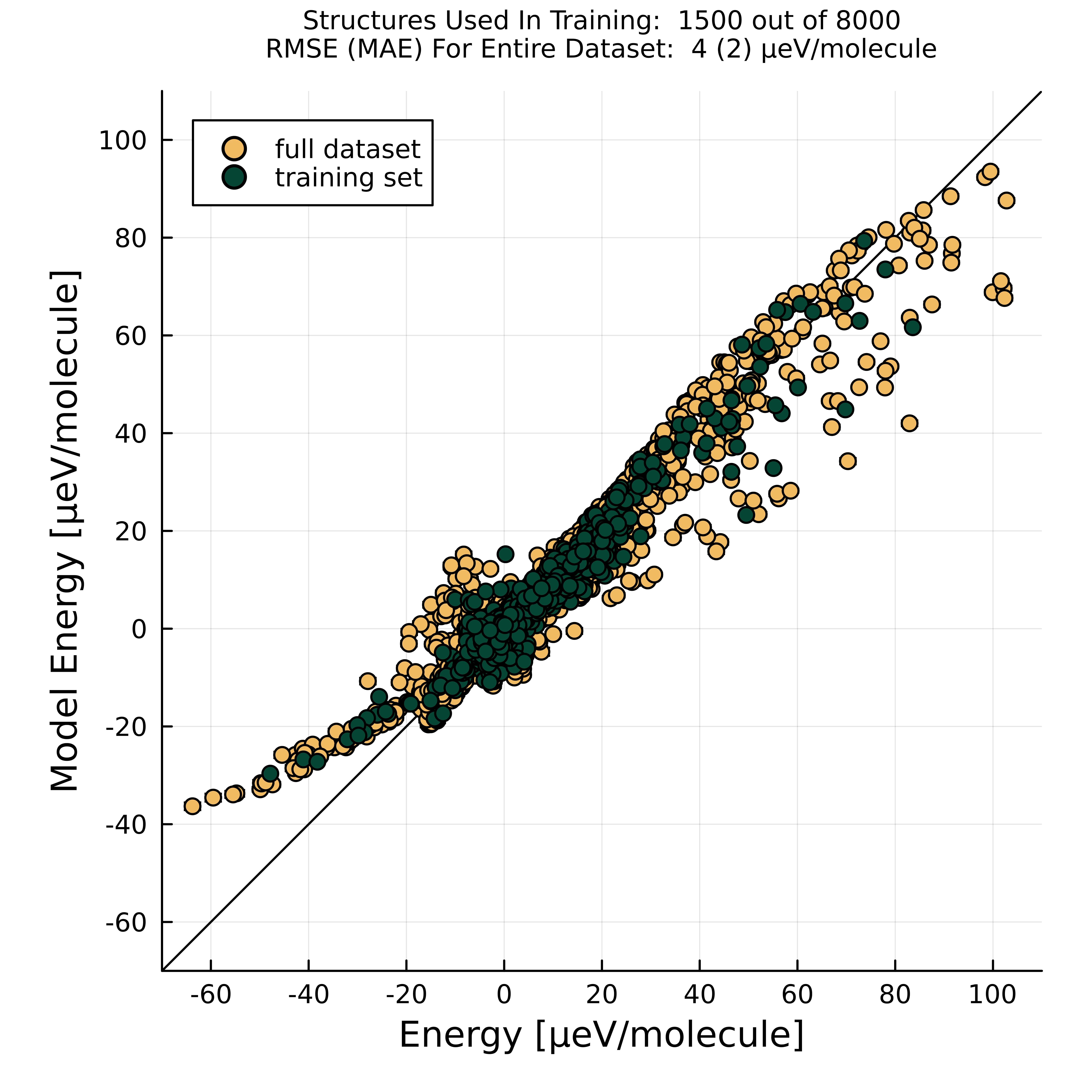} &
        \includegraphics[width=0.225\linewidth,height=\textheight,keepaspectratio]{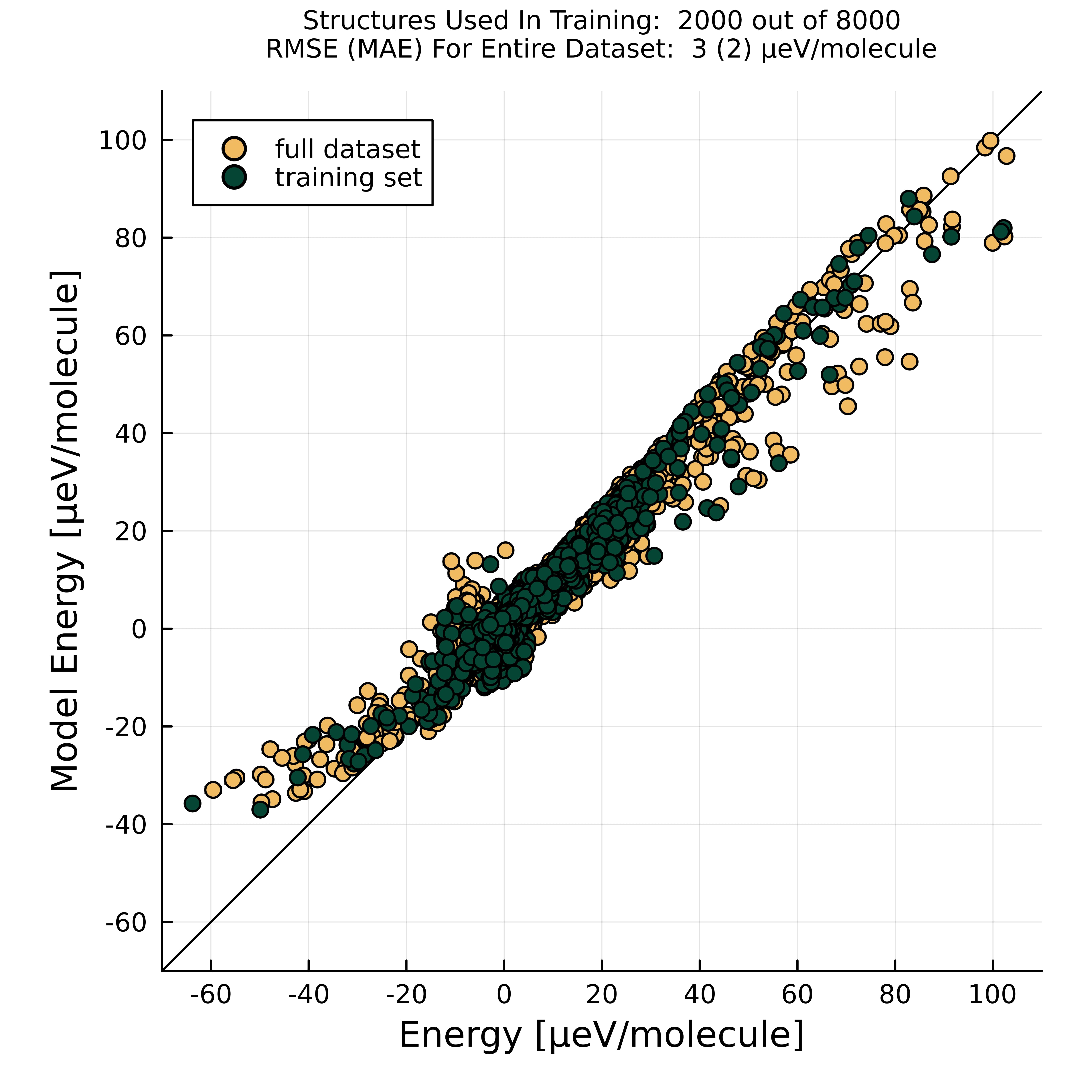}  \\
        (e) & (f) & (g) & (h)\\[6pt]
    \end{tabular}
    \begin{tabular}{cccc}
        \includegraphics[width=0.225\linewidth,height=\textheight,keepaspectratio]{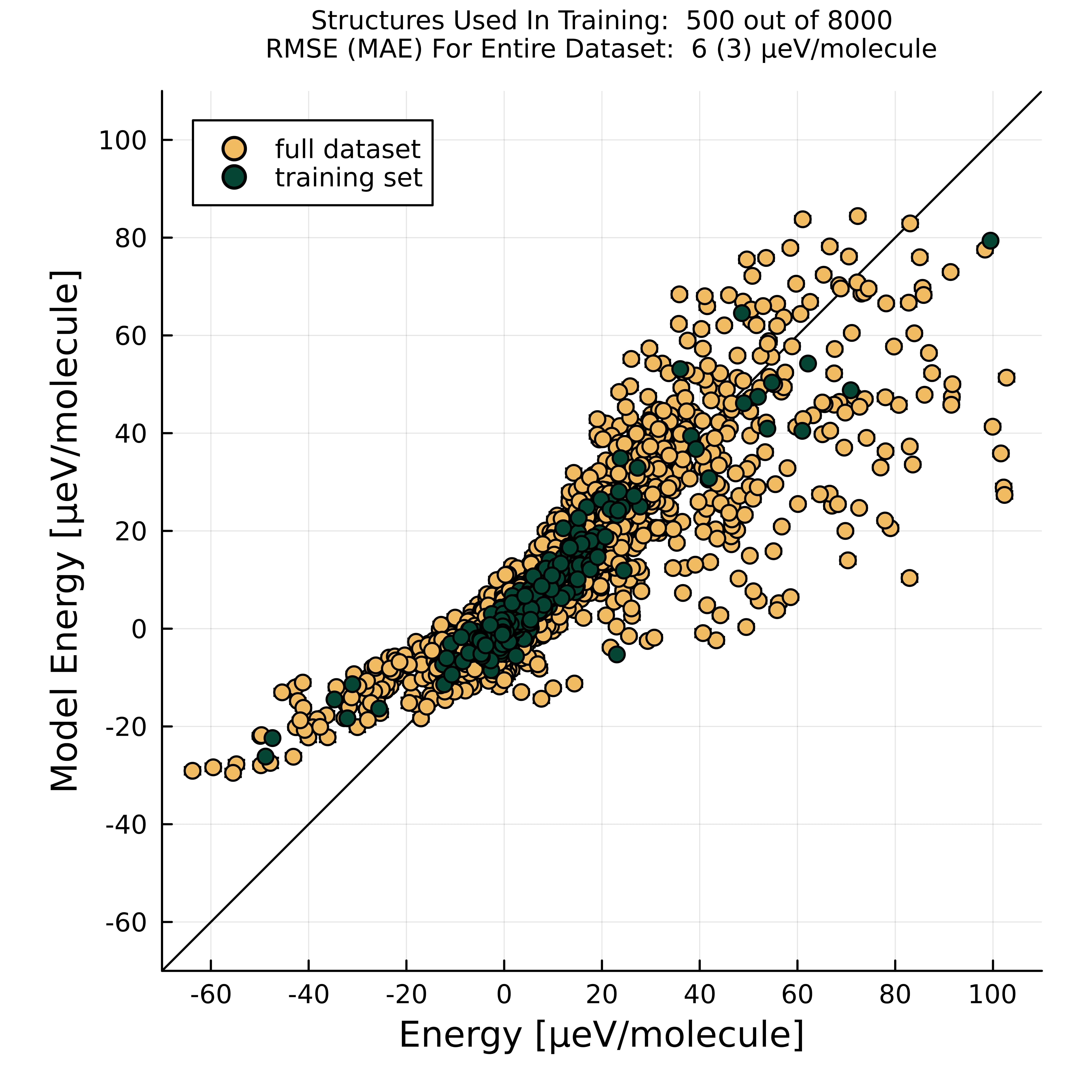} &
        \includegraphics[width=0.225\linewidth,height=\textheight,keepaspectratio]{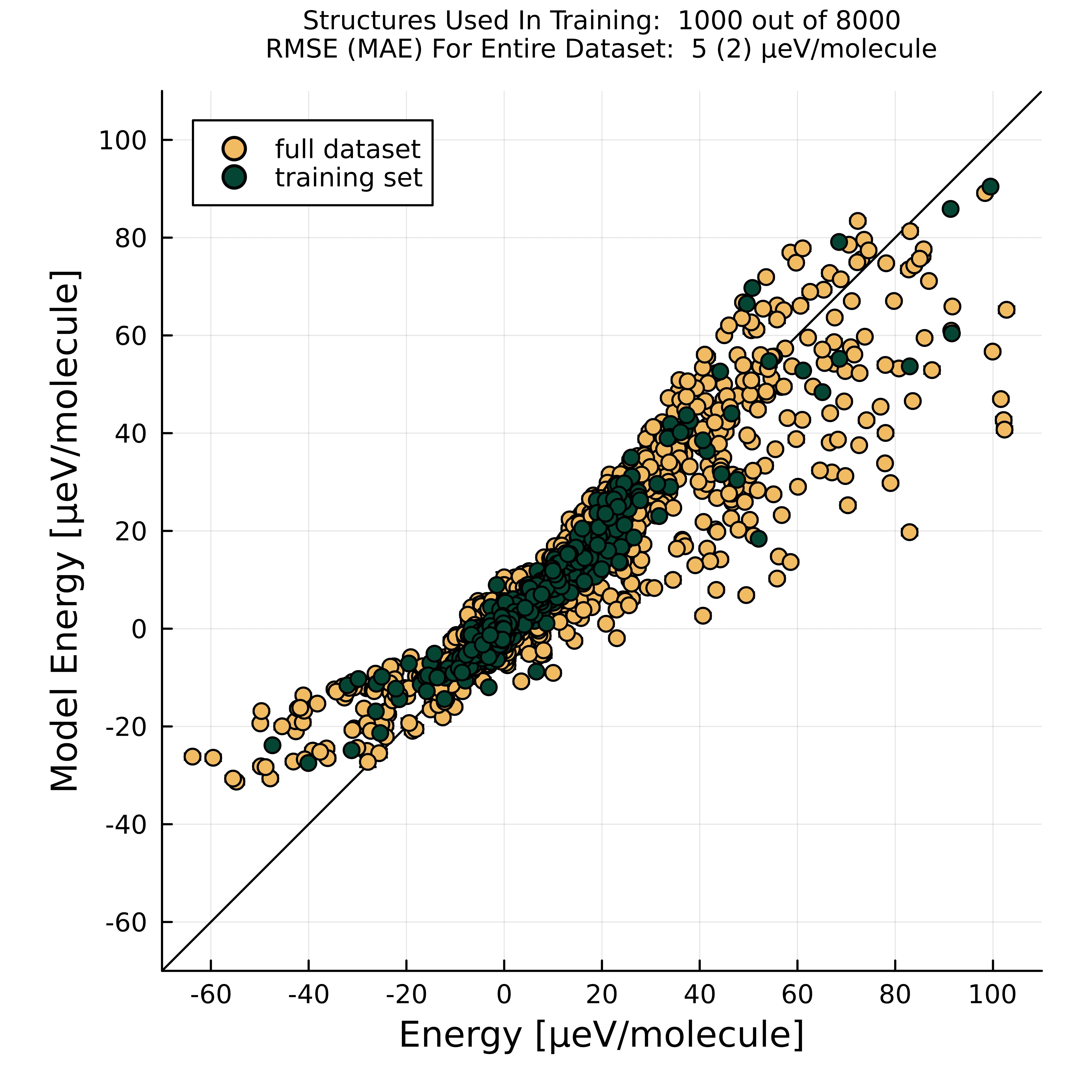} &
        \includegraphics[width=0.225\linewidth,height=\textheight,keepaspectratio]{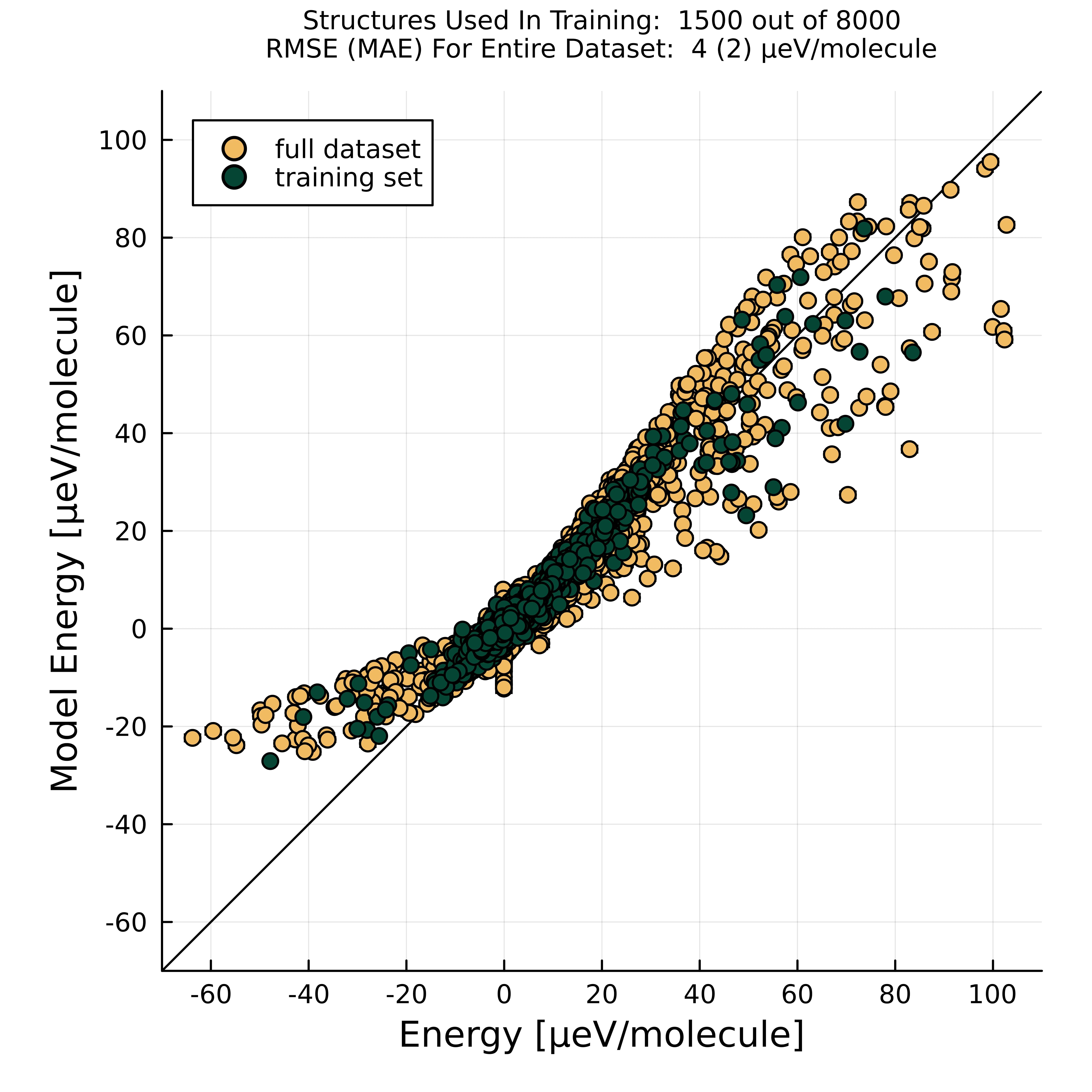} &
        \includegraphics[width=0.225\linewidth,height=\textheight,keepaspectratio]{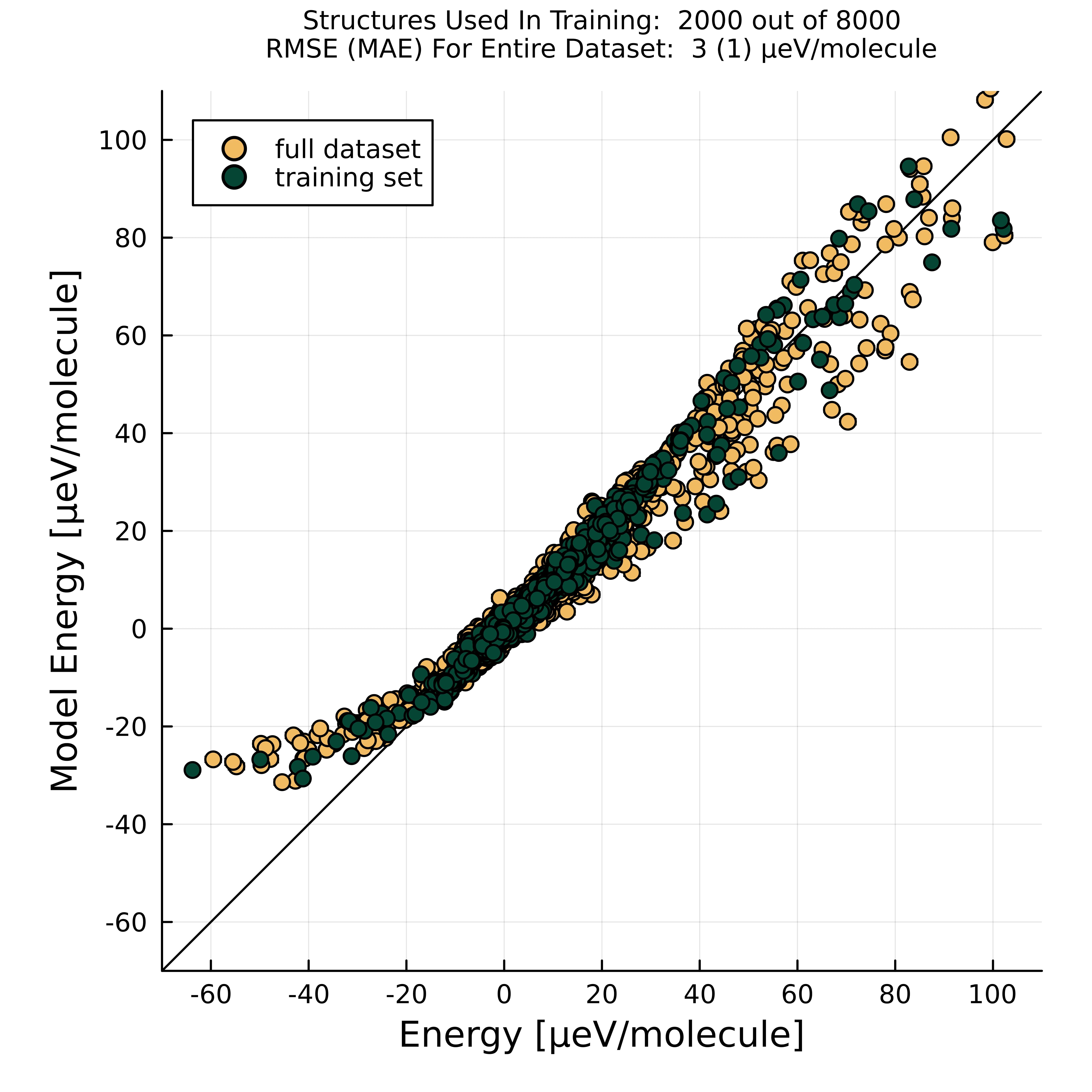} \\
        (i) & (j) & (k) & (l)\\[6pt]

    \end{tabular}
    \caption 
    {\label{fig:7}
    The plots comparing predicted and reference energies for ACE models trained on 500, 1,000, 1,500 and 2,000 structures out of a total of 8,000 dimer configurations. (a)--(d) Model trained using the \textit{Heavy-atom} representation for an \textit{ethylene} dimer. For a \textit{thiophene} system, (e)--(h) Model trained using the \textit{Heavy-atom} carbon-only representation, while (i)--(l) Model trained using the carbon-sulphur representation.}
\end{figure*}

We begin by fitting the ACE model to transfer integrals of the \textit{ethylene} and \textit{thiophene} dimer. 

The ACE models, constructed using the parameters which are mentioned in previous section, were trained on 500, 1,000, 1,500 and 2,000 dimer configurations out of a total of 8,000. 
The resulting fits using a \textit{heavy-atom} representation for \textit{ethylene} are presented in Figures~\ref{fig:7}a (500),~\ref{fig:7}b (1,000),~\ref{fig:7}c (1,500), and~\ref{fig:7}d (2,000).
With a dataset of 1,500--2,000, the ACE model can predict the \textit{ethylene} transfer integrals, with a root-mean-square error (RMSE) of 3.2~$\mu$eV/molecule and a mean absolute error (MAE) of 2.3~$\mu$eV/molecule.

The thiophene dimer is shown in Figure \ref{fig:7}e--\ref{fig:7}l.
Panels (e)--(h) show the fit from using only a purely carbon representation (Figure~\ref{fig:2}b), while Panels (i)–-(l) show the a model with separate carbon and sulphur channels (Figure~\ref{fig:2}c).
Both models are shown trained on 500, 1,000, 1,500 and 2,000 structures out of a total of 8,000. 
Now that the model understands the orientation of the molecule, stronger performance is seen. 
The RMSEs (MAEs) of 
4.95 (2.71), 3.76 (2.26), 3.62 (2.14), and 3.10 (2.03)~$\mu$eV/molecule for the carbon-only representation (Fig.~\ref{fig:2}b). 
The richer carbon-sulphur representation (Fig.~\ref{fig:2}c) has lower RMSEs (MAEs) of 5.93 (3.17), 4.67 (2.41), 3.95 (2.04) and 2.65 (1.40)~$\mu$eV/molecule. 
These fits are well below the accuracy of any underlying transfer integral calculator method. 

Comparing the two models, the inclusion of sulphur reduces the model error by $\approx15~\%$ for the RMSE and $\approx40~\%$ for the MAE, at 2,000 training structures. 
However, this modification increases the computational cost associated with evaluating the model descriptors (the basis set) and the time required for fitting the ACE model: the training wall-clock time rises from 15 to 20 minutes (2,000 training structures).

\begin{figure*}[t]\centering
\captionsetup{justification=centering}

    \begin{tabular}{ccc}
        \includegraphics[width=0.25\linewidth,height=\textheight,keepaspectratio]{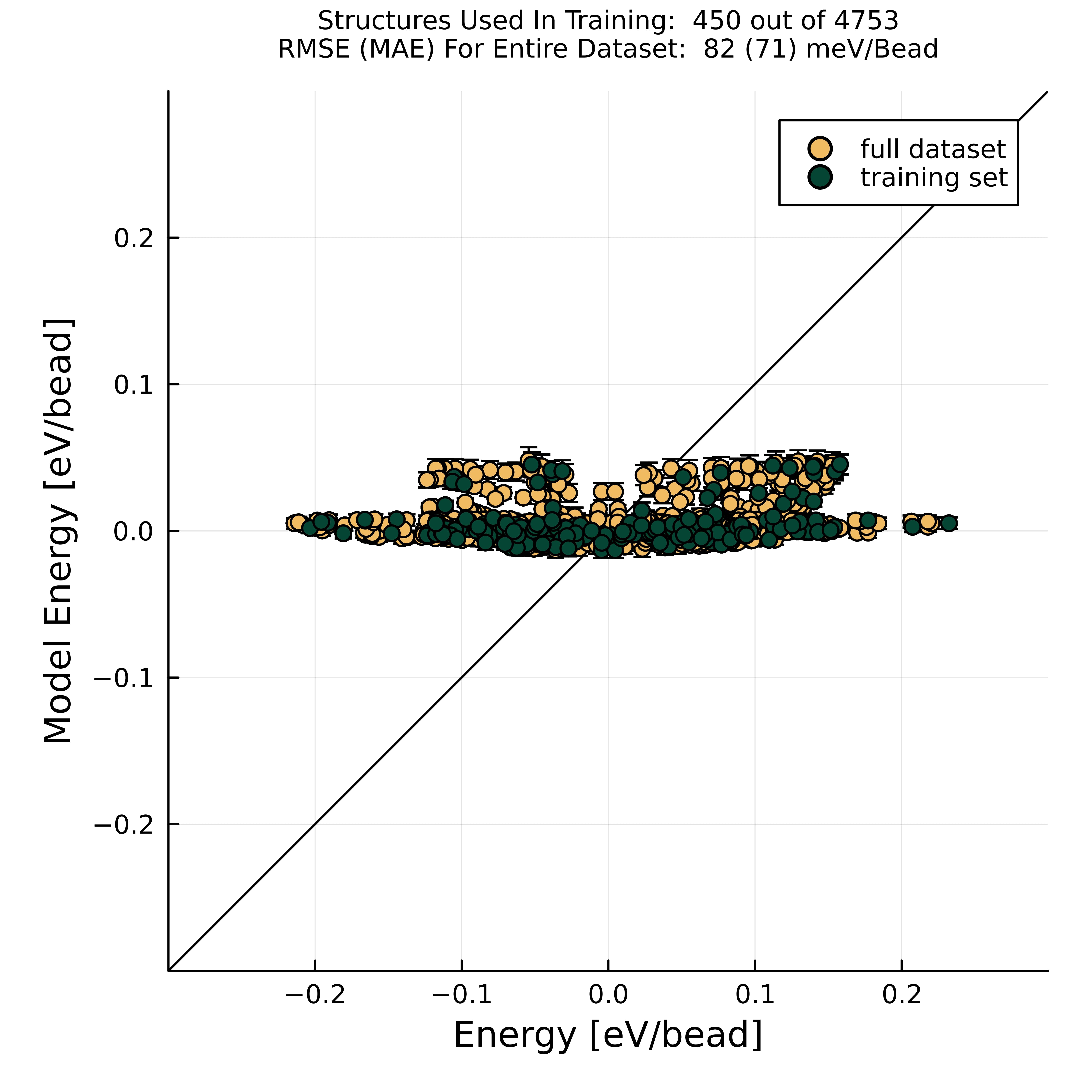} &
        \includegraphics[width=0.25\linewidth,height=\textheight,keepaspectratio]{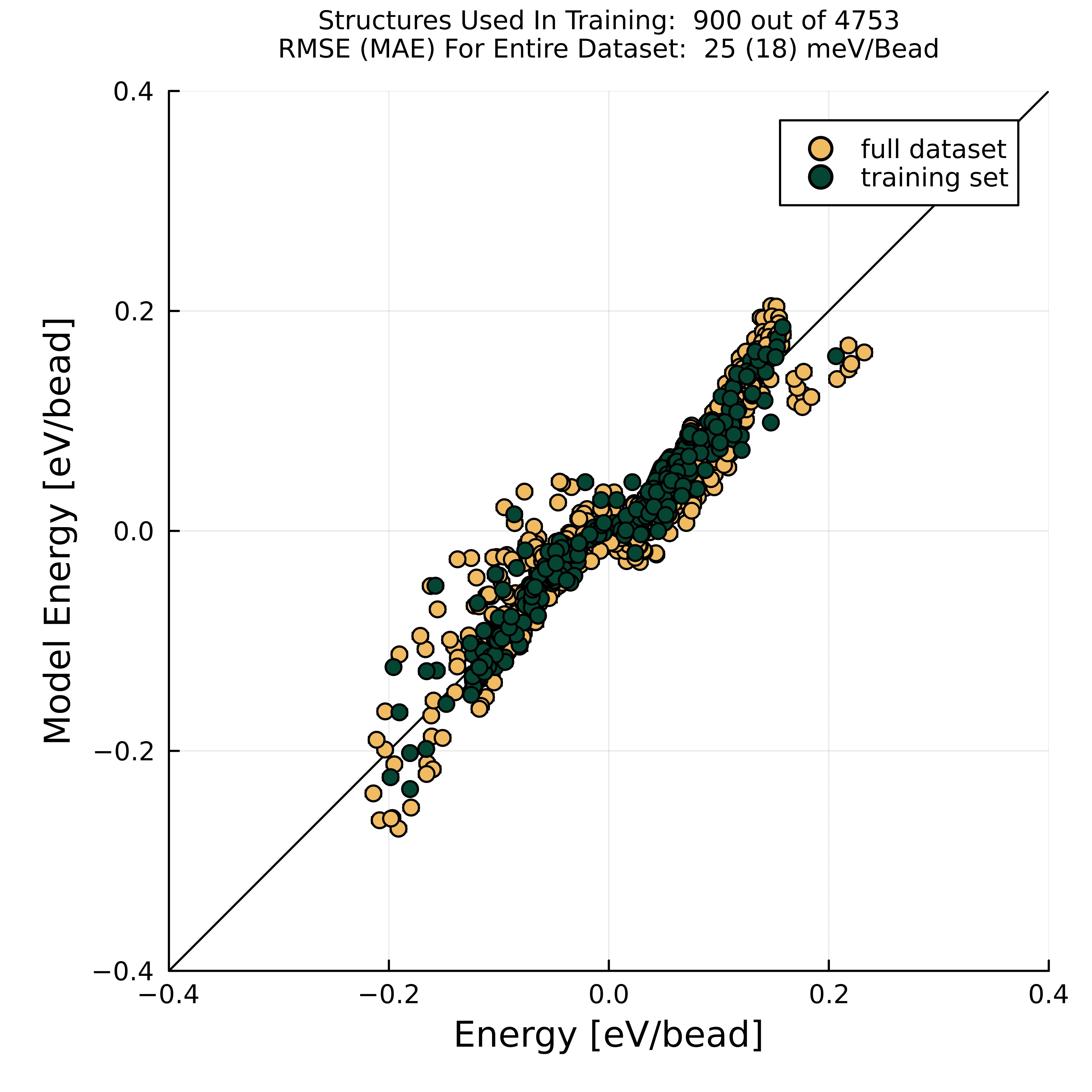} &
        \includegraphics[width=0.25\linewidth,height=\textheight,keepaspectratio]{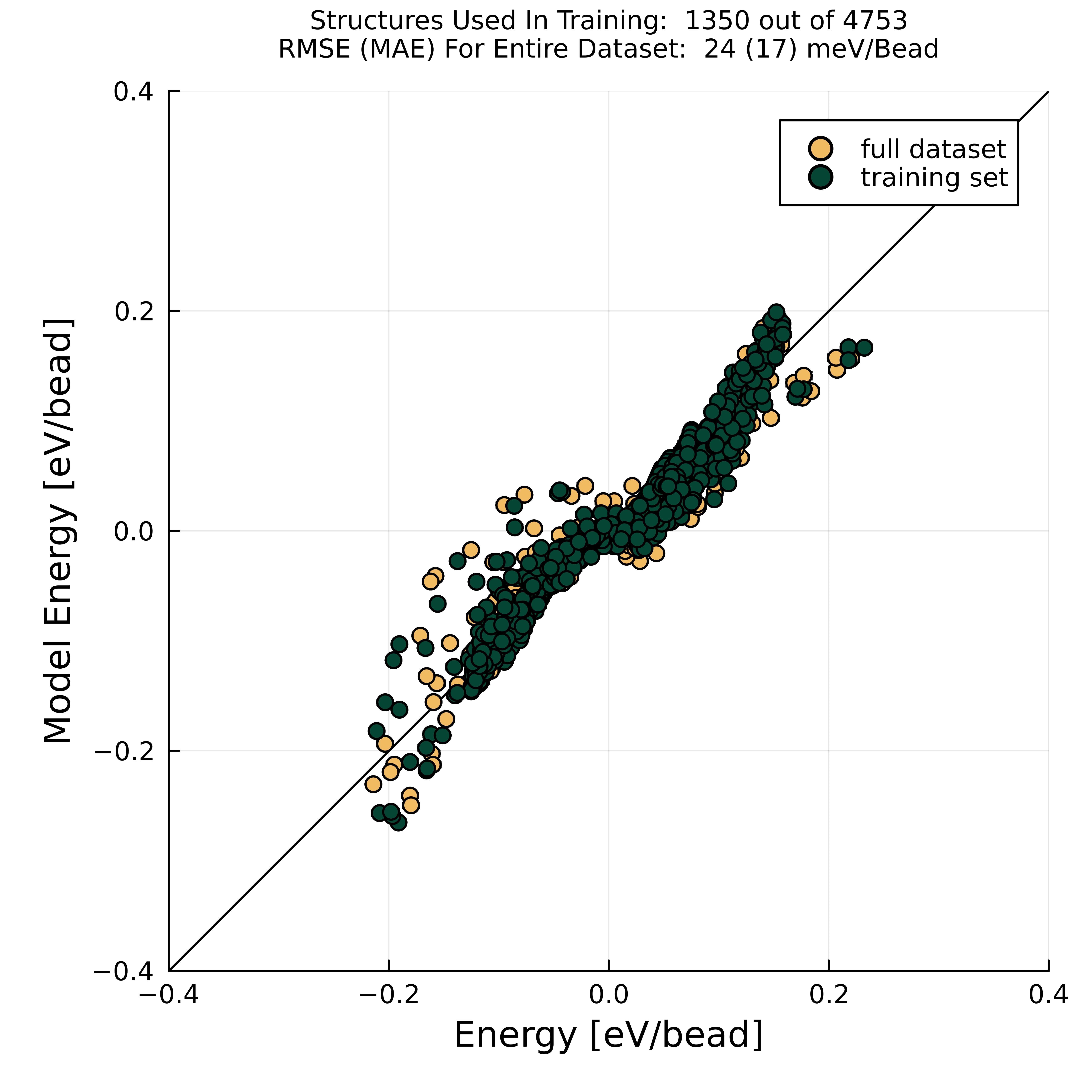} \\
        (a) & (b) & (c)\\[6pt]
    \end{tabular}
    \begin{tabular}{ccc}
        \includegraphics[width=0.25\linewidth,height=\textheight,keepaspectratio]{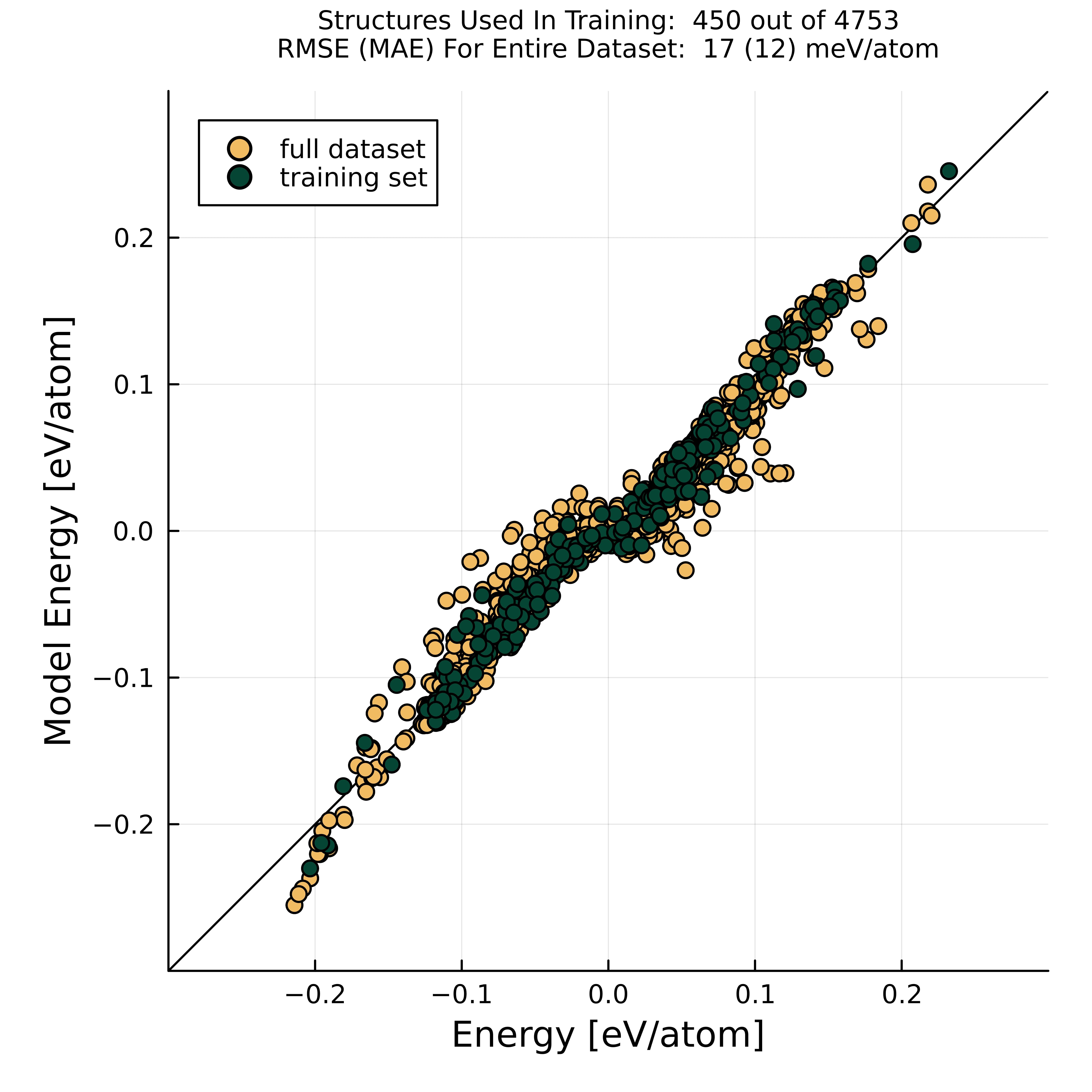} &     \includegraphics[width=0.25\linewidth,height=\textheight,keepaspectratio]{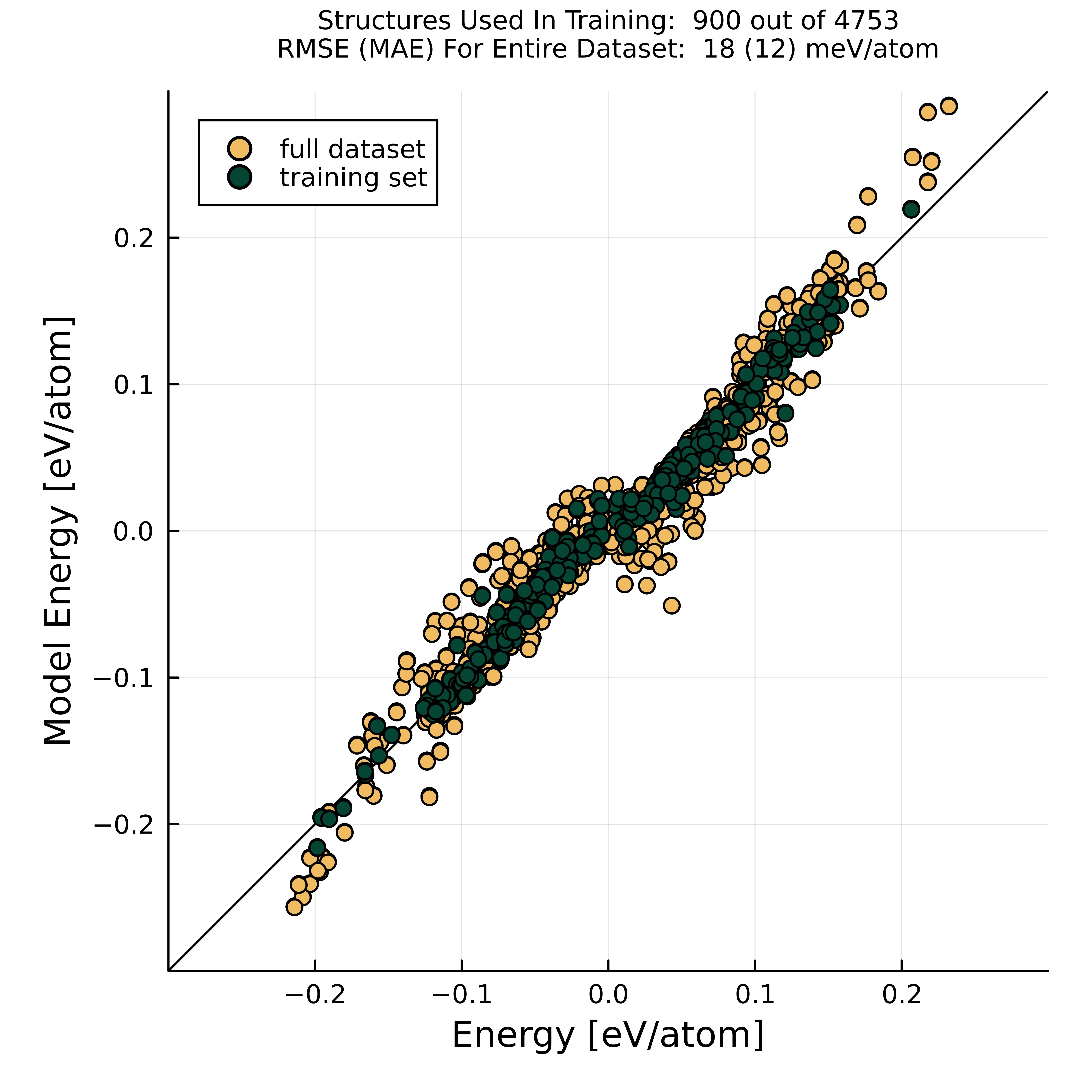} &
        \includegraphics[width=0.25\linewidth,height=\textheight,keepaspectratio]{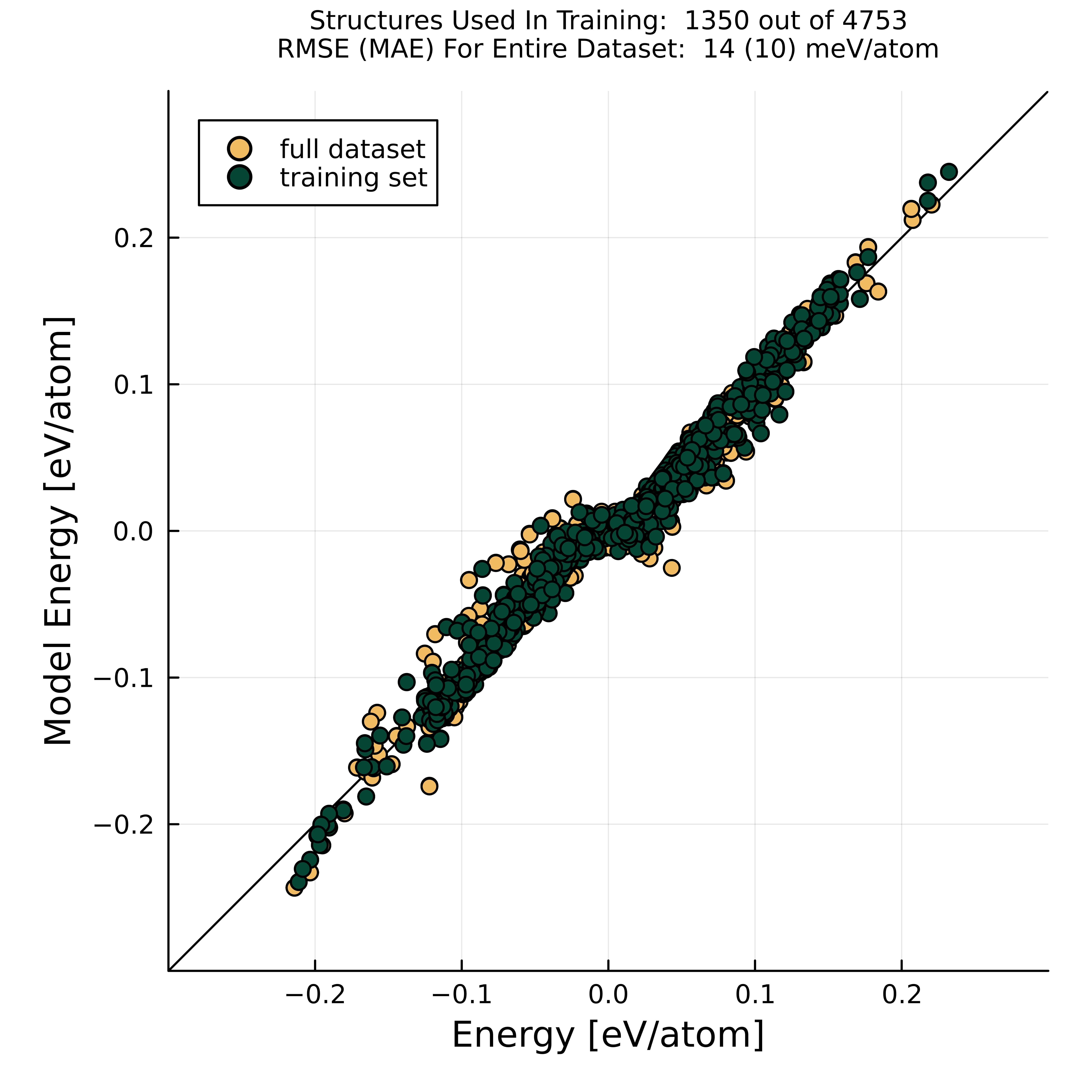}  \\
        (d) & (e) & (f)\\[6pt]
    \end{tabular}
    \caption 
    {\label{fig:8} The plots comparing predicted and reference energies for ACE models trained on 450, 900, and 1,350 structures out of a total of 4,753 \textit{naphthlene} dimer configurations. (a)-(c) Model trained using the \textit{Bead} representation. (d)-(f) Model trained using the \textit{Heavy-atom} representation.}
\end{figure*}

Finally we model \textit{naphthalene} with a two-bead coarse-grained representation, and a heavy-atom (carbon) representaton. 
With the hyper-parameters as defined previously, these ACE model were trained on 450, 900, and 1,350 dimer structures out of a total of 4,753 datapoints. 
The resulting fits using the \textit{Bead} and \textit{Heavy-atom} representations are shown in Figures~\ref{fig:8}a--c and~\ref{fig:8}e--f, respectively.

Bead-based model yields root-mean-square errors (RMSEs) of 82.36, 25.26, and 23.85~meV/molecule 
and mean absolute errors (MAEs) of 70.69, 17.93, and 16.86~meV/molecule 
In comparison, the Heavy-atom model achieves lower errors: RMSEs = 17.22, 17.75, and 13.63~meV/molecule and MAEs = 12.15, 12.36, and 9.84~meV/molecule.

Clearly our bead-model is not a good inductive bias: perhaps as a two-site model it has lost information about the relevant orientation of the two dimers. 
It appears that the more simple heavy-atom representation gives superior performance. 
Perhaps the origin of the atomic orbitals is the natural representation for an ACE model of transfer integrals. 
The heavy-atom model can fit naphthalene's transfer integral efficiently only with 450 training structures, while the bead-based model requires 900 training structures to achieve similar predictive error.
However, the cost for fitting these models is roughly identical.
Therefore, we recommend using a heavy-atom representation on similar sized molecules: our bead-representation coarse-graining would potentially offer superior performance in larger molecules (such as chromophores taken from semiconducting polymers, or fused-ring electron acceptors).

We examined our model's performance in a representative range of semiconducting molecules. 
The error of the napthalene model is adequate for charge-transfer simulations, with errors $< 2\%$ (9.84~meV/molecule).  
The error for ethylene and thiophene is significantly higher (5–10\%) when training on 2,000 samples, suggesting we need improved inductive biases for these systems. 
We hypothesise that this reduced performance is due to the inability of the model to infer the orientation in the case of ethylene (heavy-atoms), and the differing chemistry of the $C_2$ and $C_3$ carbons in thiophene (which we force the model to smear together). 
Clearly there is plenty of useful work to do in experimenting with fitting the more complex organic electronic molecules of modern technical relevance, and with different representations for their structure. 

To contextualise our findings, we compare to the literature. 
There are limitations in this comparison, as we have fitted different datasets. 
The development of standard benchmarks for organic electronic transfer integrals would significantly aid the community in the development and validation of these methods, and make them useful tools for modelling real materials. 

We highlight the data efficiency of our method. 
Recent research\cite{Lin2023,Wang2020} indicates mean absolute errors (MAEs) of 1.99 meV for naphthalene and 0.27 meV for ethylene utilising multi-stage kernel ridge regression (KRR) models with 15,000-–40,000 training samples \cite{Lin2023,Wang2020}, and approximately 6.5 meV for naphthalene employing artificial neural networks (ANNs) trained on 120,000 samples \cite{Wang2020}. 
Our naphthalene model attains a similar accuracy with 1,350 samples. 

\section{Conclusion}
We have established and validated a machine learning framework utilising the Atomic Cluster Expansion (ACE) for accurately predicting intermolecular transfer integrals in organic semiconductors.  Using ACE descriptors that integrate radial and angular information, while imposing suitable symmetry constraints, we developed models adept at understanding the complicated geometric associations of electronic couplings.

We implemented customised molecular representations, including heavy-atom and coarse-grained (Bead) descriptors, to optimise model accuracy and computing performance.  Our results demonstrate that ACE models trained on a minimum of 1,500 configurations can attain mean absolute errors $<4$~$\mu$eV/molecules for ethylene and thiophene dimers. Moreover, for the naphthalene dimer, both bead-based and heavy-atom models works efficiently and yield $<4$~meV/molecule prediction errors.


The results confirm the potential of ACE as a reliable surrogate model for charge transport simulations in organic molecular materials. 
Our approach offers state of the art data efficiency, with reliable fitting. 
The most simple heavy-atom representation was the most powerful approach we found, which may be due to the natural analogy of an ACE model and the linear combination of atomic orbitals (LCAO) used in transfer integral and quantum-chemical calculations. 
We hope that further development of these methods, undertaken hand-in-hand with applying the methods to more complex materials of direct technical relevance for organic electronics, will lead to further improvements in data efficiency and performance. 
This would then enable much higher-throughput simulations of organic electronic materials. 


\section{Author contribution}
K.K.: 
Formal Analysis (lead); 
Investigation (lead); 
Methodology (equal);
Software (lead);
Writing – original draft (equal); 

C.O.: 
Methodology (equal);
Writing – original draft (editing);

J.M.F.: 
Conceptualization (lead); 
Methodology (equal);
Writing – original draft (equal); 

\section{Acknowledgement}
J.M.F. is supported by a Royal Society University Research Fellowship
(URF-R1-191292). 
K.K. is supported by a Thai scholarship, Development and Promotion of Science and Technology project. 
Julia\cite{Julia} codes implementing these calculations are available as a repository on GitHub 
%

\onecolumngrid 
\appendix 

\section{Transfer Integrals from Atomic Cluster Expansions}

\bibliography{GaussianProcessPotentialEnergySurface}

\end{document}